\documentclass{JHEP3}
\usepackage{amsmath,amsfonts,graphicx}

\newcommand{\be}        {\begin{equation}}
\newcommand{\bea}       {\begin{equation}\begin{array}l}
\newcommand{\cbb}       {\mathbb{C}}
\newcommand{\ee}        {\end{equation}}
\newcommand{\eear}      {\end{array}}
\newcommand{\eps}       {\varepsilon}
\newcommand{\erf}[1]    {(\ref{#1})}
\newcommand{\hcal}      {\mathcal{H}}
\newcommand{\Hh}        {{\sf H}}
\newcommand{\labl}[1]   {\label{#1}\ee}
\newcommand{\lieg}      {{\sf g}}
\newcommand{\lieh}      {{\sf h}}
\newcommand{\nop}       {{\bf :}}
\newcommand{\om}        {\omega}
\newcommand{\one}       {{\bf 1}}
\newcommand{\rbb}       {\mathbb{R}}
\newcommand{\Sh}        {{\sf S}}
\newcommand{\Tr}        {{\rm Tr}}
\newcommand{\Wc}        {\mathcal{W}}
\newcommand{\Wh}        {{\sf W}}
\newcommand{\zbb}       {\mathbb{Z}}
\newcommand{\zcft}      {Z_{\rm cft}}
\newcommand{\zmm}       {Z_{\rm mm}}

\title{Higher order loop equations \\ 
for $A_r$ and $D_r$ quiver matrix models}

\author{Stefano Chiantese${}^{1)}$, 
        Albrecht Klemm${}^{2)}$, 
	Ingo Runkel${}^{1)}$\\
	${}^{1)}$ Institut f\"ur Physik, Humboldt-Univ.\ Berlin,
	   Newtonstr.\ 15, 12489 Berlin, Germany\\
        ${}^{2)}$ UW-Madison Physics Dept, 
  5211 Chamberlin Hall, 
  Madison, WI 53706-1390,
  USA\\
        E-mail: \email{chiantes, ingo@physik.hu-berlin.de}, 
	        \email{aklemm@physics.wisc.edu}
       }

\preprint{\hepth{0311258}\\ HU-EP-03/76}

\abstract{
  We use free boson techniques to investigate $A$-$D$-$E$-quiver
  matrix models. Certain higher spin fields in the free boson
  formulation give rise to higher order loop equations valid
  at finite $N$. These fields form a special kind of $\Wc$-algebra,
  called Casimir algebra. We compute explicitly the loop equations
  for $A_r$ and $D_r$ quiver models and check that at large $N$ 
  they are related to a deformation of the corresponding singular
  Calabi-Yau geometry.
}


\begin{document}

\section{Introduction}

Since the work of Kontsevich \cite{Kontsevich:ti} it is well appreciated 
that matrix model techniques are very effective in solving topological 
string theory questions. Recently Dijkgraaf and Vafa proposed a new 
matrix model description for the topological string on non-compact Calabi-Yau 
threefolds, which are ALE fibrations of $A$-$D$-$E$ type over the complex plane
\cite{Dijkgraaf:2002vw}. Various checks have been made on the claim that 
the matrix model solves the closed topological string on these backgrounds 
\cite{Klemm:2002pa, Dijkgraaf:2002yn} and gives the 
exact holomorphic information of 
$\mathcal{N}{=}1$ quiver gauge theories, namely the superpotential and the gauge kinetic 
terms \cite{Dijkgraaf:2002pp}. Orbifold and orientifold constructions 
widen the range of geometries and the field theories questions that can be addressed, 
see e.g. 
\cite{Feng:2002gb,Klemm:2003cy,Landsteiner:2003rh,Argurio:2003ym} 
and references therein.

The loop equations encode the symmetries of the matrix model as 
Ward identities, which are in principal strong enough to 
determine the correlators. Moreover the geometry of the local Calabi-Yau 
manifold is directly described by the large $N$ limit of the loop 
equations for the resolvent, which encodes the spectral 
density of the matrix. 

One of the most elegant methods to obtain these equations is to 
rewrite the matrix model correlators in terms of CFT correlators 
in a theory of free chiral bosons. Correlators of products of 
bosonic currents are identified with matrix model correlators 
involving polynomials of the resolvent. Particular combinations of 
normal ordered products of the bosonic currents can be systematically 
identified with so-called Casimir fields. Their correlators fulfill a vanishing 
condition, which translates in the desired loop equations. 
This method is well known \cite{Kharchev:1992iv, Morozov:hh, Kostov:1999xi} 
and it was used in the context of the matrix model for the topological B-model 
to establish those terms in the loop equations of the $A$-quiver matrix models, 
which lead to the identification with the $A$ singularities in the fibre of 
the local Calabi-Yau space \cite{Dijkgraaf:2002vw}.  
The Casimir fields form a $\Wc$-algebra, which is the symmetry algebra
of the matrix model \cite{Dijkgraaf:1990rs, Fukuma:1990jw}. 
As opposed to a normal Lie algebra, the commutator
of two modes in a $\Wc$-algebra does not necessarily give only a finite sum of modes,
but can also result in an infinite sum of products of modes.
The special $\Wc$-algebra formed by the Casimir fields has been
called a Casimir algebra \cite{Bais:1987dc, Bouwknegt:1992wg}.

In the present paper we want to formulate the approach precisely 
enough to derive explicitly the exact finite $N$ loop equations 
for $A_r$ and $D_r$ quiver theories, which are our main result 
and are given in equations \erf{eq:Ar-quiv-loop} and 
\erf{eq:D-r-loop}, \erf{eq:D-loop}.
The techniques of section \ref{sec:ADE-FB} equally apply to the $E$-series,
but we do not work out loop equations explicitly for this case.
The large $N$ loop equations enable us to identify 
the $A_r$ and $D_r$ fibre geometries together with the subleading 
terms, which encode the resolution of the singularity by 
renormalisable deformations, i.e.\ complex structure moduli of 
the topological string B-model. 

These data calculate the exact $\mathcal{N}{=}1$ gauge theory information. 
The finite $N$ loop equations encode the exact terms of the gauge theory 
coupled to gravity. In the topological string context they 
encode the $g>0$ genus correlators, which was checked for 
$A_2$ using the formalism of \cite{Akemann:1996zr}
in \cite{Klemm:2002pa, Dijkgraaf:2002yn} for genus 1.
The equivalence between the generalised Konishi anomalies of the 
$\mathcal{N}{=}1$  theory and the loop equations, 
observed in the large $N$-limit in \cite{Cachazo:2002ry}, 
was generalised to the gravitational sector in \cite{David:2003ke,Alday:2003ms}.

The occurrence of the chiral CFT in the derivation of the 
loop equation seems auxiliary at the first glance. However 
it is known that the Virasoro constraints, which naturally
appear in the CFT formulation, characterise topological string
correlation functions for topological gravity, see 
\cite{Dijkgraaf:1991qh} for review. The higher spin 
fields used to obtain the higher loop equations should lead to 
similar constraints for the topological string theory correlators 
in the quiver backgrounds. Recent results 
\cite{Aganagic:2003db} for the open topological string established  
a chiral boson on a Riemann surface, which encodes the essential 
part of the Calabi-Yau geometry, as actually describing the 
complex structure deformation of the topological B-model. The Ward 
identities for the chiral boson correlators are 
strong enough to solve the all genus open string amplitudes for 
the topological vertex. Open/closed string duality extends this logic 
to the closed topological string. One should find a CFT formulation, 
which is identified with Kodaira-Spencer gravity for the closed string,  
and use the $\Wc$-constraints to directly fix the all genus partition 
function of the topological string on the quiver geometries.              

Another motivation to study quiver matrix models comes from 
\cite{Alexandrov:2003pj}.
There it is argued that $\tau$-functions should be thought of as ``the 
next generation of special functions'', whose properties should be
investigated on their own right and which will be relevant to describe
partition functions of string theory.
This program was started in \cite{Alexandrov:2003pj} 
with the $\tau$-function given
by the hermitian one matrix model, and among the obvious next
candidates would be the quiver matrix models.

{\bf Acknowledgements.} We would like to thank V.~Kazakov, I.~Kostov and
C.~Schweigert for helpful discussions. I.R.\ thanks the SPhT Saclay for
hospitality, where part of this work was completed.
S.C.\ is supported by DFG Graduiertenkolleg 271/3--02 and
I.R.\ is supported by the DFG project KL1070/2--1.

\section{A-D-E quiver matrix models and free bosons}\label{sec:ADE-FB}

\subsection{Loop equations and matrix models}

Let us summarise the method we will use in the derivation
of the quiver loop equations in the well-studied case of
the hermitian one-matrix model, 
see e.g.\ the review \cite{DiFrancesco:1993nw}.
The model is defined by the matrix integral
\be
   \zmm ~=~ ({\rm const}) \int\!\!d\Phi\,e^{-\frac{1}{g_s} \Tr W(\Phi) }
   ~= \int_{-\infty}^\infty \!\!\!\!\! d\lambda_1 \dots d\lambda_N
    \!\!\! \underset{m<n}{\prod_{m,n=1}^N} \!\!(\lambda_m{-}\lambda_n)^2 \, 
    e^{-\frac{1}{g_s} \sum_{k=1}^N W(\lambda_k)}
\labl{eq:herm-mat}
The first integral is over all hermitian $N{\times}N$ matrices,
while the second integral amounts to expressing the first one in
terms of the eigenvalues of $\Phi$.
In the notation used below, this is the $A_1$-quiver model
with the real axis as integration contour.
The observables of the matrix model are traces of powers
of $\Phi$,
\be
  \zmm\big[\,\Tr(\Phi^{m_1}) \cdots \Tr(\Phi^{m_n})\,\big]
  ~=~ ({\rm const}) \int \!\! d\Phi \,
  \Tr(\Phi^{m_1}) \cdots \Tr(\Phi^{m_n}) \,
  e^{-\frac{1}{g_s} \Tr W(\Phi) } ~~.
\labl{eq:herm-mat-corr}
These correlators can be conveniently expressed in
terms of a generating function
\be
  \zmm\big[\, \om(z_1) \cdots \om(z_n) \,\big] ~,
\labl{eq:herm-mat-corr-gen}
where we introduced the resolvent $\om(z)$ of the matrix model,
which is defined as
\be
  \om(z) ~=~ \frac{1}{N} \, \Tr \frac{1}{z-\Phi} ~=~ 
  \frac{1}{N} \sum_{k=1}^N \frac{1}{z-\lambda_k} ~~.
\ee
Expanding \erf{eq:herm-mat-corr-gen} in powers of $z_k$
gives as coefficients the correlators \erf{eq:herm-mat-corr}.
The resolvent obeys the so-called loop equation, obtained
by a change of variables $\Phi \rightarrow \Phi + \eps /(z{-}\Phi)$
in the integral \erf{eq:herm-mat},
\be
  \zmm\big[\, \om(z)^2 - \frac{1}{N g_s} W'(z) \om(z) 
  + p(z) \,\big] =0 ~~,
\labl{eq:herm-mat-loop}
where $p(z)$ is a polynomial of maximal degree equal to twice the
degree of $W'(z)$. The loop equation determines the expectation value
of the resolvent in the large $N$ limit, and it can also be used to 
set up an iterative procedure to find its $1/N$-expansion 
\cite{Ambjorn:1992gw,Akemann:1996zr}.

There is an alternative way to derive the loop equation
\erf{eq:herm-mat-loop} by relating the matrix model to the
conformal field theory of one free boson in two dimensions
via $\zcft = \frac{1}{N!} \zmm$
\cite{Kharchev:1992iv, Morozov:hh, Kostov:1999xi}.
Here $\zcft$ is
a shorthand for the vacuum correlator of the free boson in a complicated
background which encodes the matrix integral. This relation is further
used to map CFT correlators of $n$-fold normal ordered products of
the free boson current $J(z)$ to matrix model correlators of polynomials in 
the resolvent,
\be
  \zcft\big[\, \nop J(z)\cdots J(z) \nop \,\big] ~=~ ({\rm const'}) \,
  \zmm\big[ \,\big( \om(z) - \frac{1}{2Ng_s} W'(z) \big)^n \,\big] ~~.
\labl{eq:intro-rel1}
To recover the loop equations one needs in addition the relation
\be
  \oint_{\mathcal{C}} \frac{dz}{2\pi i} ~ 
  \frac{1}{z{-}x} ~
  \zcft\big[\, T(z) \,\big] ~=~ 0 ~~,
\labl{eq:intro-rel2}
where $T(z) = \frac12 \nop J(z)J(z) \nop$ is the holomorphic stress tensor 
of the free boson and $\mathcal{C}$ is an integration contour
specified in section \ref{sec:cas2loop}. Also in this section,
it is explained how \erf{eq:intro-rel1} and \erf{eq:intro-rel2}
can be derived from the explicit form of $\zcft$. From 
\erf{eq:intro-rel2} one finds that $\zcft\big[T(z)\big]$ is regular
on all of $\cbb$, and in fact one can show that it is equal to
a polynomial $\tilde p(z)$ of degree less or equal to twice the 
degree of $W'(z)$. Combining with \erf{eq:intro-rel1} one finds
\be
   \tilde p(z) = \frac12 \zcft\big[\, \nop J(z) J(z) \nop \,\big]
   = ({\rm const''})
   \,\zmm\big[\, \big(\om(z) - 
   \frac{1}{2Ng_s} W'(z)\big)^2 \,\big] ~~,
\ee
which is the same as the loop equation \erf{eq:herm-mat-loop} 
after expressing $\tilde p(z)$ in terms of $p(z)$.

For the quiver matrix models, this procedure
is generalised in several ways. The quiver models are multi-matrix
models (section \ref{sec:ADE-def}) and accordingly 
we have not only one, but several
resolvents $\om_i(z)$. Further, these resolvents obey not only a 
loop equation of second order in the $\om_i(z)$, but there is also a set
of higher order loop equations. 
Attempting to find these equations by change of variables 
in the matrix integrals turns out to be extremely tedious,
and exploiting the relation to CFT proves 
a far more efficient method. 
The CFT now consists of several free bosons
(section \ref{sec:freeboson}),
and not only the stress tensor leads to a loop equation, but there
are also currents of spin higher than two, which 
obey a relation analogous to \erf{eq:intro-rel2} and give 
corresponding higher order loop equations
(sections \ref{sec:higher} and \ref{sec:cas2loop}).

These loop equations are not only essential for solving the matrix
model, but describe also the background geometry of the type
IIB string,  whose low energy limit is the $\mathcal{N}{=}1$ ADE quiver
gauge theory \cite{Dijkgraaf:2002vw}, see section \ref{sec:Ar-large-N}
and \ref{sec:Dr-large-N} for the identification
of the corresponding geometries from the large $N$ loop equations.

\subsection{The A-D-E quiver matrix model}\label{sec:ADE-def}

In general, a quiver matrix integral of rank $r$ takes the form
\cite{Kostov:1992ie, Kostov:1995xw, Dijkgraaf:2002vw}
\be
  \zmm = ({\rm const}) \int \prod_{i=1}^r d\Phi_i \prod_{\langle m,n \rangle}
  dQ_{mn} ~ e^{ -\frac 1{g_s} \Tr W(\Phi,Q) } ~.
\labl{eq:quiver-int}
Here the $\Phi_i$ are $N_i \times N_i$ matrices and the 
$Q_{mn}$ are $N_m \times N_n$ matrices. 
The potential $W(\Phi,Q)$ is given by
\be
  W(\Phi,Q) ~= \sum_{i,j=1}^r s_{ij} \, Q_{ij} \, \Phi_j \, Q_{ji}
  + \sum_{i=1}^r W_i(\Phi_i) ~,
\labl{eq:quiver-pot}
for some polynomials $W_i(x)$. The constants $s_{ij}$ are
antisymmetric, $s_{ij} = - s_{ji}$, 
they obey $s_{ij} = 1$ if $i<j$ and the nodes in the quiver
diagram are linked, and $s_{ij}=0$ otherwise.
The notation $\langle m,n \rangle$ for the range of the product
in \erf{eq:quiver-int} denotes all pairs $(m,n)$ with $1\le m,n \le r$
s.t.\ $s_{mn} \neq 0$. The definition of the integration region
requires some care, see also \cite{Lazaroiu:2003vh}.
We will address this problem in section \ref{sec:contour}.

In principle one can proceed by integrating out
the $Q_{mn}$ and expressing $\zmm$ as an integral over 
the eigenvalues $\lambda_{i,I}$ of $\Phi_i$, where $I=1,\dots,N_i$. 
This would result in the expression
\be
  \zmm = \int \prod_{k,K} d\lambda_{k,K} ~ 
  \prod_{i=1}^r \underset{I<J}{\prod_{I,J=1}^{N_i}} 
  (\lambda_{i,I} - \lambda_{i,J})^2 
  ~\underset{i<j}{\prod_{i,j=1}^r} ~ \prod_{I=1}^{N_i} \prod_{J=1}^{N_j} 
  \,(\lambda_{i,I} - \lambda_{j,J})^{-|s_{ij}|} 
  ~e^{ -\Sh } ~,
\labl{eq:quiver-ev-int}
where
\be
  \Sh = \frac 1{g_s} \sum_{l,L} W_l(\lambda_{l,L}) ~,
\labl{eq:quiver-action}
and the constant in \erf{eq:quiver-int} 
has to be chosen appropriately. In practice this would require
a consistent definition of the integration regions for the
$\Phi_i$ and $Q_{mn}$ to avoid divergences. For the purpose of
this paper we will take \erf{eq:quiver-ev-int} -- or rather a
regularised form thereof, described in section \ref{sec:contour} 
-- as a definition
of the matrix model and think of \erf{eq:quiver-int} as a motivation
for considering an integral of the form \erf{eq:quiver-ev-int}.

{}From the general quiver models one obtains the A-D-E matrix models
by choosing the $|s_{ij}|$ to take the special form
\be
  |s_{ij}| = 2 \delta_{ij} - A_{ij} ~,
\ee
where $A_{ij}$ is the Cartan matrix of a rank $r$ Lie algebra of
A-D-E type. In this case we have $A_{ij} = (\alpha_i,\alpha_j)$, where
the $\alpha_i$ are the simple roots. The integral \erf{eq:quiver-ev-int}
can now be written more compactly as
\cite{Kostov:1992ie, Kharchev:1992iv, Morozov:hh, 
Kostov:1995xw, Dijkgraaf:2002vw}
\be
  \zmm = \int \prod_{k,K} d\lambda_{k,K} 
  \underbrace{\prod_{i,j=1}^r \prod_{I=1}^{N_i} \prod_{J=1}^{N_j}
    }_{(i,I)<(j,J)}
  (\lambda_{i,I}-\lambda_{j,J}
  )^{(\alpha_i,\alpha_j)} 
  ~ e^{ -\Sh} ~,
\labl{eq:ADE-quiv-mat}
where one has to choose an ordering of the pairs $(i,I)$.
Here we have chosen $(i,I)<(j,J)$ if either $i<j$ or else if $i=j$ and
$I<J$. Different choices of ordering change equation
\erf{eq:ADE-quiv-mat} by a factor of $\pm 1$.

\subsection{A free boson representation}\label{sec:freeboson}

In this section we will construct the integral
\erf{eq:ADE-quiv-mat} as a correlator in the
chiral CFT of $r$ free bosons 
\cite{Kharchev:1992iv, Morozov:hh, Kostov:1999xi}. 
To fix conventions,
consider one free boson $\varphi(z)$ for a start. 

The $U(1)$-current $i \partial \varphi(z)$
will be denoted by $J(z)$. The vertex operators 
corresponding to the chiral part of the exponentials 
$:\!e^{iq\varphi(z)}\!:$ of the
free boson field are called $V_q(z)$. They have $U(1)$-charge
$q$ and conformal weight $h=\frac12 q^2$. The OPEs involving
$J(z)$ read
\be
  J(z)J(w) = \frac{1}{(z{-}w)^2} + {\rm reg}(z{-}w)
  ~ , ~~
  J(z)V_q(w) = \frac{q}{z{-}w} V_q(w) + {\rm reg}(z{-}w) 
\labl{eq:fb-OPEs}
and the conformal block with $n$ insertions on the 
complex plane, together with a charge $q$ placed at
infinity, is given by
\be
  \langle q | V_{q_1}(z_1) \cdots V_{q_n}(z_n) |0\rangle
  ~=~ \delta_{q_1 + \cdots + q_n,q} 
  \prod_{i,j=1\,;\,i<j}^n (z_i-z_j)^{q_i q_j} ~.
\labl{eq:fb-block}
Here $\langle q|$ is an out-state dual to the
highest weight state $|q\rangle$ of $U(1)$-charge $q$.
By definition we have
\be
  \langle q | V_{q_1}(z_1) \cdots V_{q_n}(z_n) |0\rangle
  ~=~ \lim_{L\rightarrow\infty} L^{q^2} ~
  \langle 0 | V_{-q}(L) V_{q_1}(z_1) \cdots V_{q_n}(z_n) |0\rangle
  ~.
\ee
The modes of the current $J(z)$ are defined via
\be
  J(z) = \sum_{k\in\zbb} J_k \, z^{-k-1}
  \quad , \quad
  J_k = \oint_{\gamma_0} \frac{dz}{2\pi i} \,z^k \,J(z) ~,
\ee
where $\gamma_0$ is a circular contour around the origin.
The OPEs \erf{eq:fb-OPEs} imply the commutation
relations
\be
 [J_m,J_n] = m \,\delta_{m+n,0} 
 \quad , \quad
 [J_m,V_q(z)] = q z^m \,V_q(z) ~.
\labl{eq:J-com}
Using these it follows that for $k\ge 0$
\be
  J_k V_{q_1}(z_1) \cdots V_{q_n}(z_n)|0\rangle
  ~=~ (\sum_{i=1}^n q_i (z_i)^k) ~ 
  V_{q_1}(z_1) \cdots V_{q_n}(z_n)|0\rangle ~,
\ee
and thus also, for any out-state $\langle X|$ and
$k \ge 0$,
\be
   \langle X| e^{-t J_k}
   V_{q_1}(z_1) \cdots V_{q_n}(z_n)|0\rangle
   ~=~ e^{-t \sum_{i=1}^n q_i (z_i)^k} ~
   \langle X| 
   V_{q_1}(z_1) \cdots V_{q_n}(z_n)|0\rangle ~.
\labl{eq:eHk-in-corr}

Let us now return to the general case. Denote by $\lieg$ a
simply laced Lie algebra of rank $r$. Let $K$ be the Killing
form on $\lieg$ and fix once and for all a basis $H^a$,
$a=1,\dots,r$ of the Cartan subalgebra $\lieh$ of $\lieg$ 
s.t.\ $K(H^a,H^b) = \delta_{a,b}$. Let $\lieh^*$ be the
dual of $\lieh$ and denote by $e_a$ the basis dual to
$H^a$. Another basis of $\lieh^*$ is provided by the
$r$ simple roots $\alpha_i$ of $\lieg$. 
The bilinear form on $\lieh^*$ induced by $K$ is
denoted by $(\cdot,\cdot)$.

Consider the chiral CFT consisting of the product
of $r$ free bosons, with components $J^{(a)}(z)$ and
$V^{(a)}_q(z)$, where $a=1,\dots,r$. 
To an element $u = \sum_a u^a e_a$ of $\lieh^*$ assign
the current $J^u(z)$ and the vertex operator $V_u(z)$ as
\be
  J^u(z) = \sum_{a=1}^r u_a \, J^{(a)}(z)
  \quad {\rm and} \quad
  V_u(z) = \prod_{a=1}^r V_{u_a}^{(a)}(z) ~.
\labl{eq:rfb-conv}
One computes the conformal weight of $V_u(z)$ to be
$h(V_u) = \frac12 (u,u).$
With the conventions \erf{eq:rfb-conv}, the analogues of formulas
\erf{eq:fb-OPEs}, \erf{eq:fb-block} and \erf{eq:J-com} are given by
\bea
  \displaystyle
  J^u(z)J^v(w) = \frac{(u,v)}{(z{-}w)^2} + {\rm reg}(z{-}w)
  ~ , ~~
  J^u(z)V_q(w) = \frac{(u,q)}{z{-}w} V_q(w) + {\rm reg}(z{-}w) 
  \\[6pt]\displaystyle
  \langle q | V_{q_1}(z_1) \cdots V_{q_n}(z_n) |0\rangle
  ~=~ \delta_{q_1 + \cdots + q_n,q} 
  \!\! \prod_{i,j=1\,;\,i<j}^n \!\! (z_i-z_j)^{(q_i,q_j)} ~.
  \\[13pt]\displaystyle
  [J^u_m,J^v_n] = (u,v) \, m \, \delta_{m+n,0} \quad, \qquad
  [J^u_m,V_q(z)] = (u,q) \, z^m \, V_q(z) ~.
\eear\labl{eq:r-fb-block}
Here $u,v,q$ as well as $q_1,\dots,q_n$ are elements of
$\lieh^*$. The modes of $J^u(z)$ are 
written as $J^u_k$.

To the $r$ potentials $W_i(x)$ in \erf{eq:quiver-pot}
we assign a $\lieh^*$-valued function $\Wh(x)$ as
follows. Suppose
\be
  W_i(x) = \sum_{k=0}^\infty t^{(i)}_k x^k
  \qquad {\rm for}~i=1,\dots,r
\labl{eq:Wi-poly}
and define vectors $\tau_k \in \lieh^*$ via
$(\tau_k,\alpha_i) = t^{(i)}_k$. We then set
$\Wh(x) = \sum_{k\ge 0} \tau_k \, x^k \in \lieh^*$.
This potential enters the definition of 
an operator $\Hh$, which encodes the
potentials of the matrix model in the
free boson representation,
\be
  \Hh = \frac{1}{g_s} \sum_{a=1}^r \oint_{\gamma_0}
  \frac{dz}{2\pi i}  \, \big( \Wh(z),e_a \big)  \, J^{(a)}(z)
  = \frac{1}{g_s} \sum_{k \ge 0} J^{\tau_k}_k ~.
\labl{eq:H-def}
As a final ingredient we need the screening charges
\be 
  Q_i = \oint_{\gamma_i} \frac{dz}{2\pi i} V_{\alpha_i}(z) ~,
\ee
where $\gamma_i$ are some integration contours, to be
specified in section \ref{sec:contour}. 
Note that because $\lieg$ is simply
laced we have $(\alpha_i,\alpha_i)=2$ for all $i$ and
the vertex operators $V_{\alpha_i}(z)$ indeed have
conformal weight one.

The free boson representation of the quiver matrix integral 
$\zmm$ in \erf{eq:ADE-quiv-mat} is now given by 
\cite{Kharchev:1992iv, Morozov:hh, Kostov:1995xw, Kostov:1999xi}
\be
  \zcft = \langle \vec N |
  e^{-\Hh} e^{Q_1} \cdots e^{Q_r} |0\rangle 
  \qquad {\rm where} \quad
  \langle\vec N| = \langle N_1 \alpha_1{+}{\cdots}{+}N_r \alpha_r |~,
\labl{eq:fb-Z}
which is equal to $\zmm$ up to a $N_i$-dependent constant 
\be
 \zcft ~=~ C_{\vec N} \, \zmm \qquad {\rm ~~where~~} 
 C_{\vec N} = \frac{1}{N_1!\cdots N_r!} ~~.
\ee 
To see this equality note that
\bea\displaystyle
  \zcft 
  ~=~ \frac{1}{N_1!\cdots N_r!} ~
  \langle \vec N | \,
  e^{-\frac{1}{g_s} \sum_{k} J^{\tau_k}_k} \,
  (Q_1)^{N_1} \cdots (Q_r)^{N_r} \, |0\rangle
  \\[10pt] \displaystyle
  =~ \frac{1}{N_1!\cdots N_r!} ~
  \Big(\prod_{k,K} \oint_{\gamma_k} \frac{d\lambda_{k,K}}{2\pi i}\Big)
  ~ e^{-\frac{1}{g_s} \sum_{\ell} \sum_{i,I} 
  (\tau_\ell,\alpha_i) (\lambda_{i,I})^\ell} \,
  \langle \vec N | \,
  \prod_{j,J} V_{\alpha_j}(\lambda_{j,J}) \, |0\rangle ~,
\eear\ee
where in the first step we inserted the definition of $\Hh$
and used that by conservation of the $r$ $U(1)$-charges only
one term of the exponentials of the screening charges in 
\erf{eq:fb-Z} can contribute. In the second step the 
screening charges have been replaced by the corresponding
integrals and the relation \erf{eq:eHk-in-corr} has been
employed to remove the modes $J^{\tau_k}_k$.
Inserting the explicit expression for the $r$-boson
conformal blocks in \erf{eq:r-fb-block} one recovers
the integral \erf{eq:ADE-quiv-mat}.

\subsection{Integration contours}\label{sec:contour}

As already mentioned, in this paper we will consider
\erf{eq:quiver-int} as a formal expression motivating
the eigenvalue representation \erf{eq:ADE-quiv-mat},
which we will take as a definition of the A-D-E quiver matrix model.
However, even in \erf{eq:ADE-quiv-mat} the integrand contains
potential singularities, namely whenever 
$(\alpha_i,\alpha_j) < 0$ and two eigenvalues
belonging to the nodes $i$ and $j$ approach 
each other. Below, in a procedure similar to \cite{Lazaroiu:2003vh}, 
the expression \erf{eq:ADE-quiv-mat}
is defined as a limit of a regularised integral.
The regularisation involves a number of 
ad hoc choices, but these turn out not to affect the
loop equations we will derive.

\FIGURE[bt]{
\begin{picture}(300,100)(0,0)
\put(70,0){ 
  \put(0,0){\includegraphics[scale=0.5]{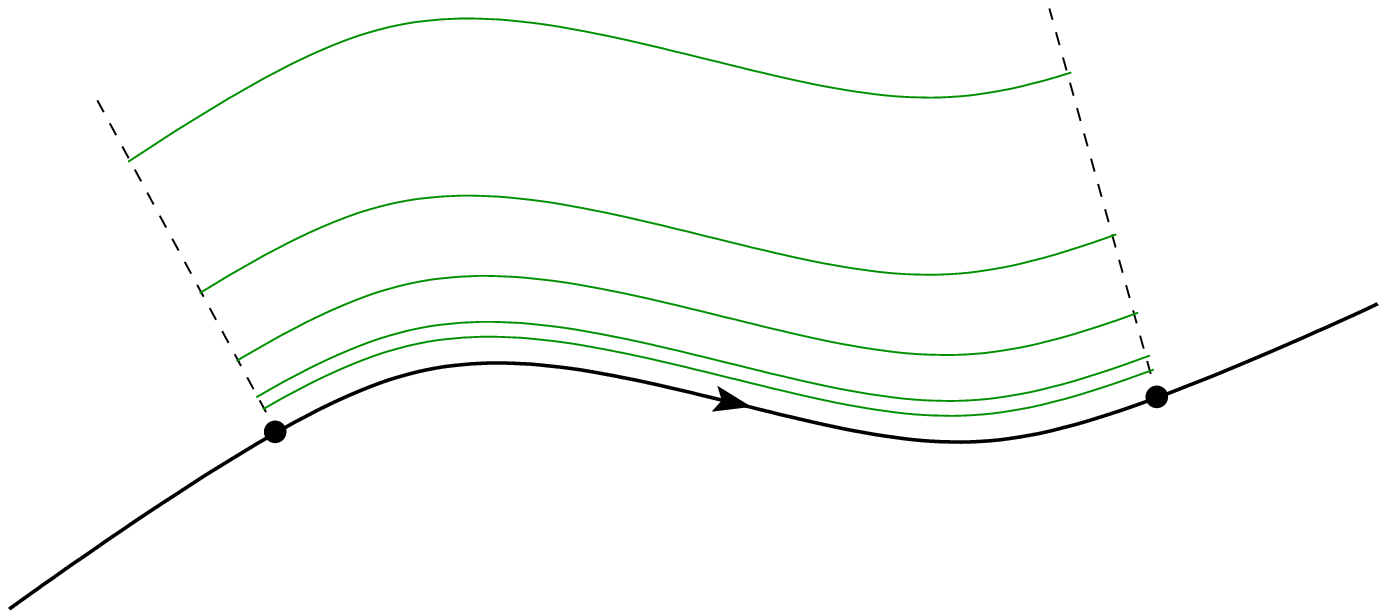}}
  \put(45,18){\scriptsize $\gamma(-L)$}
  \put(170,23){\scriptsize $\gamma(L)$}
  \put(19,63){\scriptsize $\eps$}
  \put(22,43){\scriptsize $\eps/2$}
  \put(80,90){\scriptsize $\gamma^{\eps,L}_{1,1}$}
  \put(100,60){\scriptsize $\gamma^{\eps,L}_{1,2}$}
  }
\end{picture}
\caption{The original integration contour $\gamma$ and
the shifted and truncated contours $\gamma^{\eps,L}_{k,K}$
used in the regularisation prescription.}
\label{fig:contours}
}

Let $\gamma : \rbb \rightarrow \cbb$ be a smooth contour
parametrised such that it has unit tangent $|\dot\gamma(t)|=1$.
Given a small $\eps>0$ define a family of contours
$\gamma^{\eps,L}_{k,K}$ on the interval $[L,-L]$ as
\be
  \gamma^{\eps,L}_{k,K}(t)
  = \gamma(t) + \frac{ i \, \eps }{K + \sum_{j=1}^{k-1} N_j} 
  \dot\gamma(t)~,
\ee
see figure \ref{fig:contours}.
With these contours we define a regularised integral
$I^{{\rm reg}(\eps,L)}[\,\cdot\,]$ as
\be
  I^{{\rm reg}(\eps,L)}\big[\,f(\lambda_{1,1},\dots)\,\big]
  = \Big(\prod_{k=1}^r \prod_{K=1}^{N_k} \oint_{\gamma^{\eps,L}_{k,K}} 
  \frac{d\lambda_{k,K}}{2\pi i}\Big) f(\lambda_{1,1},\dots) ~.
\labl{eq:int-op}
Here it is implied that the contours 
$\gamma^{\eps,L}_{k,K}$ are integrated on the interval $[L,-L]$
only.

Using the integral operator \erf{eq:int-op} 
one defines the regularised partition functions
$\zmm^{{\rm reg}(\eps,L)}$ and $\zcft^{{\rm reg}(\eps,L)}$.
With this choice of contour the various eigenvalues
$\lambda_{k,K}$ are always at a finite distance from each other
and $\zmm^{{\rm reg}(\eps,L)} = \zcft^{{\rm reg}(\eps,L)} / C_{\vec N}$
is a (finite) number, because it is defined
as a multiple integral over a finite region of a bounded function.

Below we will be working with
\be
  \zmm^{\rm reg} = \lim_{L\rightarrow\infty} \zmm^{{\rm reg}(\eps,L)}
\labl{eq:zreg-def}
which a priory still depends on $\eps$. However the loop equations
turn out to be independent of this parameter. 
Note also that requiring the limit \erf{eq:zreg-def} to be finite
imposes further constraints on the initial contour $\gamma$.
The regularised integral \erf{eq:ADE-quiv-mat} converges if
(but not only if) 
${\rm Re}\!\big(W_i(\gamma(L))\big) \rightarrow \infty$ 
as $L\rightarrow\pm\infty$ for all $1{\le}i{\le}r$.

\subsection{Higher spin currents commuting with screening charges}
\label{sec:higher}

The loop equations in the matrix model will be constructed
from fields in the chiral algebra, which commute with the
screening charges. These fields will be called {\em Casimir
fields} for a reason explained in section \ref{sec:casimir}.

Before turning to the definition of Casimir fields
we recall some notations to express the OPE of chiral
fields. We will use the conventions of \cite{DiFrancesco:nk}, 
section 6.5. For two chiral fields $A(z)$, $B(z)$ the
chiral fields $\{AB\}_n(z)$ are defined via
\be
  A(z) B(w) = \sum_{n=-\infty}^{n_0} \frac{ \{AB\}_n(w) }{(z{-}w)^n} ~.
\ee
The (generalised) normal ordered 
product $(AB)(z)$ of two chiral fields
is defined to be
\be
  (AB)(z) = \oint_{\gamma_z} \frac{dx}{2\pi i}
  \frac{A(x)B(z)}{x-z} = \{AB\}_0(z) ~,
\ee
where $\gamma_z$ is a circular contour winding tightly
around $z$.
Multiple normal ordered products are defined recursively, 
e.g.\ $(ABCD)(z) \equiv (A(B(CD)))(z)$. In general this form of normal
ordering is neither associative nor commutative, see
e.g.\ \cite{DiFrancesco:nk} appendix 6.C, or the appendix
of \cite{Bais:1987dc} for details. We will also be using this
generalised normal ordering for free boson fields, where it reduces
to the usual notion of moving creation operators to
the left.

Consider a general spin $s$ chiral field $W(z)$,
\be
  W(z) ~=~ \sum_{n=1}^s ~ \sum_{a_1,\dots,a_n} 
  \hspace*{-0.5em}\underset{m_1+\cdots+m_n=s-n}{\sum_{m_1,\dots,m_n \ge 0}}
  \hspace*{-1em}d^{m_1\dots m_n}_{a_1\dots a_n}
  ~\big(\partial^{m_1}J^{(a_1)}\cdots \partial^{m_n}J^{(a_n)}\big)(z) ~.
\labl{eq:W-by-J}
We will give two equivalent definitions for
$W(z)$ to be a Casimir field. 

\medskip

\noindent{\bf Definition~1:} The spin $s$ 
chiral field $W(z)$ is a Casimir field if
$V_0 W(0) |0\rangle = 0$, for all 
$V(z) = V_{\alpha_i}(z)$, $i=1,\dots,r$. Here
$V_0$ denotes the zero mode of the current $V(z)$.

\medskip

\noindent{\bf Definition~2:} $W(z)$ is a Casimir field if,
for $i=1,\dots,r$, the OPE of $V_{\alpha_i}(x)$ with $W(z)$
is a total derivative in $x$,
\be
  V_{\alpha_i}(x) W(z) = \frac{\partial}{\partial x}
  \Big( \sum_n (x-z)^n A_n(z) \Big) ~,
\labl{eq:cas-def2}
where the $A_n(z)$ are some chiral fields.

\medskip

To see that the two definitions are equivalent, first note
that, by the state-field correspondence, $V_0 W(0) |0\rangle = 0$
is equivalent to
$\oint_{\gamma_z} \frac{dx}{2\pi i} V(x) W(z) = \{VW\}_{1}(z) = 0$. 
Inserting \erf{eq:cas-def2} in the last expression, we see that definition~2 implies
definition~1. Conversely, if $\{VW\}_{1}(z) = 0$ we can write
\be
  V(x) W(z) = 
  \sum_{n=-\infty}^{n_0} \frac{ \{ VW \}_n(z) }{(x{-}z)^n}
  = \frac{\partial}{\partial x}
  \Big( \underset{n\neq 1}{\sum_{n=-\infty}^{n_0}} 
  \frac{\{ VW \}_n(z)}{(1{-}n)(x{-}z)^{n-1}}  \Big) 
\labl{eq:VW-OPE}
so that definition~1 implies definition~2.

\FIGURE[t]{
\begin{picture}(500,50)(0,0)
\put(20,0){ 
  \put(0,0){\includegraphics[scale=0.5]{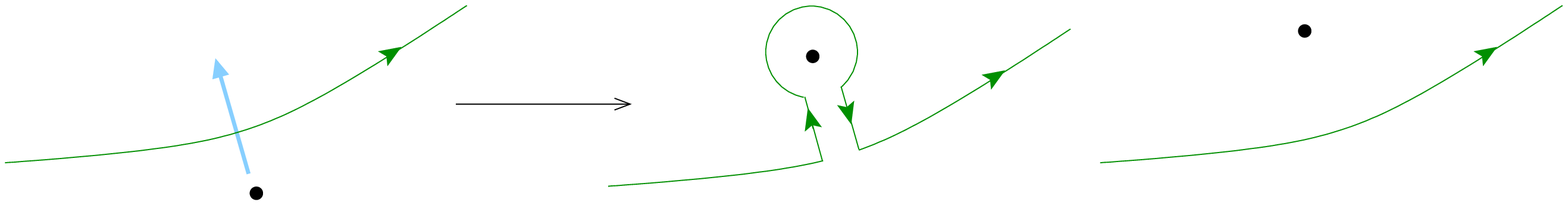}}
  \put(85,41){\scriptsize $V(x)$}
  \put(230,35){\scriptsize $V(x)$}
  \put(350,41){\scriptsize $V(x)$}
  \put(65,0){\scriptsize $W(z)$}
  \put(198,33){\scriptsize $W$}
  \put(320,40){\scriptsize $W(z)$}
  \put(260,21){$=$}
  }
\end{picture}
\caption{If a Casimir field $W(z)$ is analytically continued
through a screening charge integral, the screening integral
can be deformed back to its original form. Casimir 
fields commute with screening integrals.}
\label{fig:W-cont}
}

\FIGURE[b]{
\begin{picture}(300,110)(0,0)
\put(40,0){ 
  \put(0,0){\includegraphics[scale=0.5]{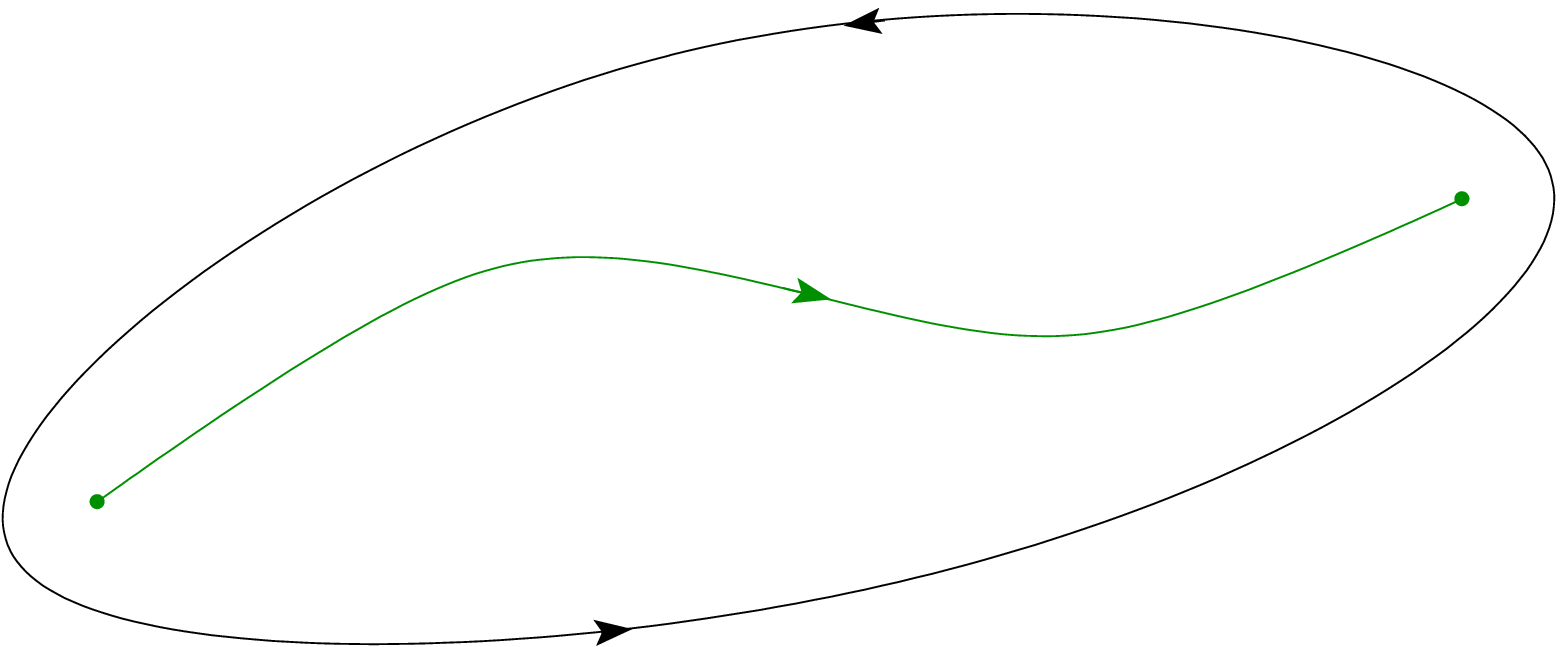}}
  \put(65,60){$\gamma_1$}
  \put(100,41){\scriptsize $V(y)$}
  \put(10,15){\scriptsize $y_0$}
  \put(205,70){\scriptsize $y_1$}
  \put(55,7){\scriptsize $f(z) W(z)$}
  \put(198,28){$\gamma_2$}
  }
\end{picture}
\caption{Integral of a Casimir field $W(z)$ times
a function $f(z)$ (contour
$\gamma_2$) around a regularised screening charge
integral with contour $\gamma_1 = \gamma^{\eps,K}_{k,K}$.}
\label{fig:double-int}
}

As we have seen while arguing the equality of the
two definitions, a Casimir field $W(z)$ has the
property $\oint_{\gamma_z} \frac{dx}{2\pi i} V(x) W(z) = 0$,
i.e.\ a contour integral of $V(x)$ around $W(z)$ can
be left out. It follows that $W(z)$ does not
have a discontinuity across the screening charge
integration contours, as illustrated in figure \ref{fig:W-cont}.
This is the first important property of Casimir fields.
The second property appears when integrating $W(z)$
around a regularised screening charge contour 
$\gamma_{k,K}^{\eps,L}$ as in figure \ref{fig:double-int},
where $\gamma_1 = \gamma_{k,K}^{\eps,L}$ and
$y_0 = \gamma_1(-L)$,
$y_1 = \gamma_1(L)$. The function $f(z)$ is assumed 
to be analytic inside the $z$-integration contour 
$\gamma_2$, and will later on be set to $f(z) = (z-x)^{-1}$ 
for some point $x$ outside the contour $\gamma_2$.
The double integration indicated in figure \ref{fig:double-int} can be
solved
\be
  \oint_{\gamma_1} \frac{dy}{2 \pi i}
  \oint_{\gamma_2} \frac{dz}{2 \pi i}
  f(z) \cdot W(z)V(y)
  = A(y_1) - A(y_0) ~,
\labl{eq:A-A}
with $A(y)$ given by
\be
  A(y) = \sum_{k=0}^{n_0-2} ~ \sum_{l=0}^{n_0-k-2}
  \frac{(-1)^{k+l} ~ \partial^l f(y)}{k! \, l! \, (k{+}l{+}1)}
  ~\partial^k \{VW\}_{k+l+2}(y) ~.
\labl{eq:A-form}
In verifying this calculation one can substitute
the OPE \erf{eq:VW-OPE}, then
Taylor-expand both $\{VW\}_n(z)$ and $f(z)$
around the point $y$ and carry out the 
$z$-integration, which gives rise to a 
Kronecker delta, removing the $n$-summation.
The $y$-integration can be solved trivially
because of the derivative $\partial/\partial y$
introduced by the OPE.

The ability \erf{eq:A-A} of $W(z)$ to replace
screening charge integrations by field insertions
at the integration boundaries is crucial in the
following argument, which ultimately leads
to the loop equations.

\subsection{From Casimir fields to loop equations}
\label{sec:cas2loop}

As a generalisation of \erf{eq:fb-Z}
let us introduce the notation
\be
  \zcft^{\rm reg}\big[\phi_1(x_1) \cdots \phi_n(x_n)\big]
  =  \langle \vec N |
  e^{-\Hh} e^{Q_1} \cdots e^{Q_r} 
  \phi_1(x_1) \cdots \phi_n(x_n) |0\rangle ~
\ee
for the (unnormalised) expectation value of some
chiral fields $\phi_k(z)$ in the matrix model
background. Consider the equation
\be
  \lim_{L\rightarrow \infty}
  \oint_{\gamma_{\rm ev}} \frac{dz}{2\pi i} ~
  \frac{1}{z{-}x} ~
  \zcft^{{\rm reg}(\eps,L)}\big[W(z)\big] = 0 ~,
\labl{eq:intW=0}
where the contour $\gamma_{\rm ev}$ surrounds
all contours $\gamma_{k,K}^{\eps,L}$, but not
the point $x$. 
This equation was
given in \cite{Dijkgraaf:1990rs, Fukuma:1990jw, Kharchev:1992iv} 
in terms of modes, and in integral form (with the $L$-limit implicit) 
in \cite{Kostov:1995xw, Kostov:1999xi} 
(for the stress tensor) as well as in \cite{Dijkgraaf:2002vw}.
Equation \erf{eq:intW=0} will be established in detail
in appendix \ref{sec:app-lim}. The outline of the argument
is as follows. The integration contour $\gamma_{\rm ev}$
is deformed to encircle each one of the regularised contours
$\gamma_{k,K}^{\eps,L}$. By \erf{eq:A-A} such an integral
can be replaced by an insertion of $A(y_1)-A(y_0)$. 
In the limit $L\rightarrow\infty$ this insertion will
cause an exponential damping in the correlator, due to
the non-trivial out-state $\langle \vec N| e^{-\Hh}$.

The rewriting of \erf{eq:intW=0} as a loop equation
will proceed in several steps. First note the
commutation relations
\be
  [J_k^{u} , \Hh^n] =  \frac 1{g_s} k \, \theta(-k) 
  ~ (u,\tau_{-k}) ~ n \, \Hh^{n-1} ~,
\ee
which can be verified by recursion. 
Here $\theta(x)$ is the Heaviside function.
As a short
calculation shows, this in turn implies that
\be
  e^{-\Hh} J^u(z) = \Big(J^u(z) -\frac 1{g_s} \big(u, \Wh'(z)\big)\Big)e^{-\Hh}
  \qquad {\rm with} \quad
  \Wh'(z) = \frac{\partial}{\partial z} \Wh(z) ~.
\labl{eq:JeH-com}
Define the positive and negative mode part of
$J^u(z)$ as
\be
  J^u_+(z) = \sum_{k \ge 0} J_k^u \, z^{-k-1}
  \quad {\rm and} \quad
  J^u_-(z) = \sum_{k < 0} J_k^u \, z^{-k-1}
\ee
so that $J^u(z) = J^u_+(z) + J^u_-(z)$.
Since $\langle \vec N| J^u_k = 0$ for all $k < 0$ it is 
easy to see that 
$\langle \vec N| J^u(z) = \langle \vec N| J^u_+(z)$. In
fact, using the definition of the normal ordered product
one can verify the relation
\be
  \langle \vec N| (\partial^{m_1}J^{u_1} \cdots \partial^{m_n}J^{u_n})(z)
  = \langle \vec N|\partial^{m_1}J^{u_1}_+(z) \cdots \partial^{m_n}J^{u_n}_+(z) ~.
\ee
Putting all this together we arrive at the conclusion
\bea\displaystyle
  \zcft^{\rm reg}\big[\,(\partial^{m_1}J^{u_1}\cdots 
  \partial^{m_1}J^{u_n})(z)\,\big]
  \\[5pt]\displaystyle
  = ~ \langle \vec N |
  \prod_{k=1}^n \Big( \partial^{m_k}J^{u_k}_+(z) - 
  \frac 1{g_s} \big(u_k, \partial^{m_k+1} \Wh(z)\big) \Big) ~
  e^{-\Hh} e^{Q_1} \cdots e^{Q_r} 
  |0\rangle ~.
\eear\labl{eq:zcft-JJ}
On the matrix model side, we have $r$ different
resolvents $\om_k(z)$, defined as
\be
  \om_k(z) = \frac 1N \sum_{K=1}^{N_k} 
  \frac{1}{z-\lambda_{k,K}}
  \qquad {\rm where} \quad 
  N = N_1 + \cdots + N_r ~.
\ee
Expectation values of the resolvents are defined
as
\bea
  \zmm^{\rm reg}\big[\,\om_{a_1}(z_1) \dots \om_{a_s}(z_s)\,\big]
 \\[1pt]  ~=~ I^{\rm reg}\Big[~
  \om_{a_1}(z_1) \dots \om_{a_s}(z_s)
  \underbrace{\prod_{i,j=1}^r \prod_{I=1}^{N_i} \prod_{J=1}^{N_j}
    }_{(i,I)<(j,J)}
  (\lambda_{i,I}-\lambda_{j,J}
  )^{(\alpha_i,\alpha_j)} 
  ~e^{ -\frac 1{g_s} \sum_{l,L} W_l(\lambda_{l,L}) } ~\Big]~.
\eear\labl{eq:zmm-ww}
The link between \erf{eq:zcft-JJ} and \erf{eq:zmm-ww}
is provided by the equations
\be
  \Big(- g_s \sum_{k\ge 0} z^{-k-1} 
  \frac{\partial}{\partial \tau_k^{(a)}} ~\Big) ~e^{-\Hh} ~=~ 
  J^{(a)}_+(z) e^{-\Hh}
\labl{eq:cft-mm-link-1}
and
\be
   \Big(- \frac{g_s}{N} \sum_{k\ge 0} z^{-k-1} 
  \frac{\partial}{\partial t_k^{(i)}} ~\Big)~ e^{-\Sh} ~=~ \om_i(z) e^{-\Sh} ~.
\labl{eq:cft-mm-link-2}
Here we have set $\tau_k^{(a)} = (\tau_k,e_a)$. Recall that
by definition $t^{(i)}_k = (\tau_k,\alpha_i) = 
\sum_{a=1}^r (\alpha_i,e_a) \tau_k^{(a)}$, which results in a corresponding
relation for the partial derivatives, namely
\be
  \frac{\partial}{\partial \tau_k^{(a)}} ~=~ 
  \sum_{i=1}^r \, (e_a,\alpha_i) ~ \frac{\partial}{\partial t_k^{(i)}} ~.
\ee
In this way one obtains, for example,
\bea\displaystyle
  \zcft^{{\rm reg}(\eps,L)}\big[J^{(a)}(z)\big] 
  =~ \Big(- g_s \sum_{k\ge 0} z^{-k-1} 
  \frac{\partial}{\partial \tau_k^{(a)}}
  ~-~ \frac{1}{g_s} \big(e_a,\Wh'(z)\big) \Big)~\zcft^{{\rm reg}(\eps,L)} 
  \\[5pt] \displaystyle
  \qquad =~ \Big( - N \sum_{i=1}^r (e_a,\alpha_i) \frac{g_s}{N}
  \sum_{k\ge 0} z^{-k-1} \frac{\partial}{\partial t_k^{(i)}}
  ~-~ \frac{1}{g_s} \big(e_a,\Wh'(z)\big)\Big)
  ~C_{\vec N} ~\zmm^{{\rm reg}(\eps,L)}
  \\[5pt] \displaystyle
  \qquad =~ C_{\vec N} ~N~ \zmm^{{\rm reg}(\eps,L)}\Big[~ 
  \sum_{i=1}^r (e_a,\alpha_i) \om_i(z)
   ~-~ \frac{1}{N g_s} \big(e_a,\Wh'(z)\big) ~\Big] ~.
\eear\labl{eq:zcft-zmm-1J}
Note, however, that in writing \erf{eq:cft-mm-link-2}
we have expanded 
$(z{-}\lambda_{k,K})^{-1} = z^{-1} \sum_{j>0} (\lambda_{k,K}/z)^{j}$.
This is possible only if $|z| > |\lambda_{k,K}|$.
For finite $L$ we can always choose a large enough $R_0(L)$ 
such that all contours $\gamma_{k,K}^{\eps,L}$ are contained in a disc
of radius $R_0(L)$ centered at the origin. From 
\erf{eq:cft-mm-link-1} and \erf{eq:cft-mm-link-2}
we thus know that \erf{eq:zcft-zmm-1J} holds
for all $z$ outside of this disc, i.e.\ $|z|>R_0(L)$.
For $|z|\le R_0(L)$ the relation then holds by analytic continuation in $z$.
Since we have established \erf{eq:zcft-zmm-1J} for all values of $L$ and $z$, it
will continue to hold if we take the limit $L\rightarrow\infty$.

In general we set, for $u\in\lieh^*$,
\be
  y_u(z) ~=~ \sum_{i=1}^r \,(u,\alpha_i)\, \om_i(z) ~-~ \frac{1}{N g_s}
  \big(u,\Wh'(z)\big) ~.
\labl{eq:y-def}
In the  $L\rightarrow\infty$ limit we then
obtain the following relation between the CFT expectation value
of a normal ordered product of free boson currents and the
matrix model expectation value of a polynomial in the resolvents:
\be
  \zcft^{\rm reg}\big[~(\partial^{m_1}J^{u_1}\cdots 
  \partial^{m_n}J^{u_n})(z)~\big]
  ~=~ C_{\vec N} \, N^n ~ \zmm^{\rm reg}\big[~\partial^{m_1}y_{u_1}(z) 
  \cdots \partial^{m_n}y_{u_n}(z)~\big] ~.
\ee
This equation, together with \erf{eq:W-by-J} and 
\erf{eq:intW=0} gives the desired relation between 
a spin $s$ Casimir field $W(z)$ and a loop equation
of maximal order $s$ in the resolvents. Concretely,
suppose we are given a Casimir field in the form
\erf{eq:W-by-J}. Then
\be
  \oint_{\gamma_{\rm ev}} \frac{dz}{2\pi i}~
  \frac{1}{z{-}x}~
  \zcft^{{\rm reg}(\eps,L)}\big[W(z)\big] ~=~ 
  C_{\vec N} \, N^s \, ( F(x) + P(x) ) ~,
\labl{eq:loop-aux-1}
where
\begin{eqnarray}
  &&F(x) = - \sum_{n=1}^s N^{n-s} \sum_{a_1,\dots,a_s} 
    \underset{m_1+\cdots+m_n=s-n}{\sum_{m_1,\dots,m_n\ge 0}}
    d_{a_1\dots a_n}^{m_1\dots m_n}
    \nonumber\\
  && \hspace{7em} \times \quad
    \zmm^{{\rm reg}(\eps,L)}\big[~\partial^{m_1}y_{(a_1)}(x)
    \cdots \partial^{m_n}y_{(a_n)}(x)~\big]
  \\[10pt] \displaystyle
  &&P(x) = \sum_{n=1}^s N^{n-s} \sum_{a_1,\dots,a_s} 
    \underset{m_1+\cdots+m_n=s-n}{\sum_{m_1,\dots,m_n\ge 0}}
    d_{a_1\dots a_n}^{m_1\dots m_n} 
    \nonumber\\
  && \hspace{7em} \times \quad
  \oint_{\gamma_\infty} \frac{dz}{2\pi i}
  \frac{1}{z{-}x}
    \zmm^{{\rm reg}(\eps,L)}\big[~\partial^{m_1}y_{(a_1)}(z)
    \cdots \partial^{m_n}y_{(a_n)}(z)~\big]
  \label{eq:F-P-def}
\end{eqnarray}
as can be seen by deforming the contour $\gamma_{\rm ev}$ into a sum of
two contours, one encircling the point $x$ and the other,
denoted by $\gamma_\infty$, a large
circle containing all the eigenvalues as well as the point $x$.
Here we also abbreviated $y_{(a)}(z) = y_{e_a}(z)$.

The function $P(x)$ is analytic in the entire complex plane. 
Furthermore the lhs of \erf{eq:loop-aux-1} behaves as $c/x$ for
some constant $c$ as $x$ tends to infinity. On the other hand,
since for the resolvents we have $\om_i(x) \cong N_i/N \cdot x^{-1}$
as $x\rightarrow \infty$ and since the potential terms with higher derivatives
are subleading in $x$, the function $F(x)$ behaves as
\be
  F(x) ~=~ \frac{(-1)^{s+1}}{(N g_s)^s} ~ \zmm^{{\rm reg}(\eps,L)}
  \sum_{a_1,\dots,a_s} 
    d_{a_1\dots a_s}^{0\dots 0}     
    \big(e_{a_1},\Wh'(x)\big) \cdots \big(e_{a_s},\Wh'(x)\big) ~+~ O(x^{sD-1})~,
\ee
where
\be
  D = \max\!\big( \deg( W'_i(x) ) ~\big|~ i = 1,\dots,r ~\big) ~.
\ee
In order for $F(x) + P(x)$ to behave as $c/x$ for $x\rightarrow\infty$,
$P(x)$ thus has to be a polynomial of maximal degree $s  D$.
Taking the $L\rightarrow\infty$ limit in \erf{eq:loop-aux-1}, 
one finally arrives at the loop equation associated to a
Casimir field in the form \erf{eq:W-by-J},
\be
  \zmm^{{\rm reg}}\!\Big[~
  \sum_{n=1}^s N^{n-s} 
  \!\! \sum_{a_1,\dots,a_s} 
  \!\! \underset{m_1+\cdots+m_n=s-n}{\sum_{m_1,\dots,m_n\ge 0}}
  \hspace*{-1.5em} d_{a_1\dots a_n}^{m_1\dots m_n} ~ 
    \partial^{m_1}y_{(a_1)}(x) 
    \cdots \partial^{m_n}y_{(a_n)}(x)
   ~~+~ P_s(x) ~ \Big] ~=~ 0 ~,
\labl{eq:cas-loop}
where $P_s(x) = -P(x) / \zmm$ is a 
polynomial of degree $\deg(P_s) \le s D$.

\subsection{Undetermined parameters in the loop equations}\label{sec:undet}

In order to use the loop equations to determine the 
matrix model correlators recursively, we need to know more
about the a priori undetermined polynomials $P_s(x)$ appearing
in \erf{eq:cas-loop}. Let us for this section
assume that all of the $W_i(x)$ have degree $D{+}1$.
Writing out the product of $y$'s in the integrand of
$P(x)$ in \erf{eq:F-P-def}, one finds the leading terms to be of
the form $(\Wh'(z))^s + \om(z) (\Wh'(z))^{s-1} + 
N^{-1} \Wh''(z) (\Wh'(z))^{s-2} + \dots$, where
we have omitted all sums, indices, etc. Because of the asymptotics
of $\om(z)$ one sees that the (a priori unknown) resolvents
start to enter $P(x)$ only at order $x^{D(s-1)-1}$, while the
coefficients of $x^{Ds}$ to $x^{D(s-1)}$ are directly determined
through the potential $\Wh(x)$.
It follows that the polynomial $P_s(x)$ entering a loop equation
of order $s$ has $D(s{-}1)$ undetermined coefficients.

On general grounds one expects an independent loop equation
of order $s$ for every Lie algebra Casimir of $\lieg$ of order $s$. 
This will be motivated in section \ref{sec:casimir}, where
the relation between Lie algebra Casimirs and a special kind of
$\Wc$--algebra, called Casimir algebras, is discussed.

The Lie algebra $\lieg$ has $r$ Casimir operators, whose
order $s$ is related to the exponents $e$ of $\lieg$ via
$s=e{+}1$. The exponents obey the relation
$\sum_{e \in \exp(\lieg)} e = R_+$, where
$R_+ = \tfrac12(\dim\,\lieg - r)$ denotes the number of
positive roots of $\lieg$. The overall number of undetermined 
coefficients in the polynomials $P_s(x)$ of the
$r$ independent loop equations now reads
\be
  \#({\rm undet.~coeff})
  = \sum_{e \in \exp(\lieg)} D(e{+}1{-}1) = D R_+ ~~.
\labl{eq:num-Ps-coeff}

The same number of free parameters appears when setting up
a large $N$-expansion. The eigenvalues of the matrices
$\Phi_i$ are located on the cuts of the resolvents $\om_i(z)$.
For small enough $g_s$
the cuts will be located around the solutions of the
classical equations of motion, which correspond to the
critical points of the quiver potential \erf{eq:quiver-pot}.
There are $D R_+$ such critical points 
\cite{Cachazo:2001gh, Cachazo:2001sg, Dijkgraaf:2002vw} 
and for the large $N$ expansion one has to fix the
corresponding $D R_+$ filling fractions. One can thus expect
a (however quite nontrivial) bijection between
the filling fractions and the parameters \erf{eq:num-Ps-coeff}.

It should be remarked that
the large $N$ expansion is in general just a formal procedure
and does not necessarily reproduce the value of the matrix 
integrals for partition function and correlators, see 
\cite{Bonnet:2000dz}.

\subsection{The quadratic loop equation as an example}

In this section we recover the familiar feature that the
quadratic loop equation is linked to the Virasoro algebra
\cite{Dijkgraaf:1990rs, Fukuma:1990jw}, 
see also e.g.\ \cite{Morozov:hh, DiFrancesco:1993nw}.

Recall that the OPE of a weight $h$ primary field $\phi(z)$
with the conformal stress tensor $T(z)$ is of the form
\be  
  T(z) \phi(w) ~=~ \frac{h}{(z{-}w)^2} \, \phi(w)
  + \frac{1}{z{-}w} \, \frac{\partial}{\partial w} \, \phi(w)
  + {\rm reg}(z{-}w) ~.
\ee
This can always be written as a total derivative in $w$
if the conformal weight $h$ is equal to one. Since all
the screening charge currents $V_{\alpha_i}(z)$ are
primary and 
have conformal weight one, $T(z)$ is a Casimir field.
In terms of the $U(1)$-currents of the $r$ free bosons
we have
\be
  T(z) = \sum_{a=1}^r \frac 12 (J^{(a)} J^{(a)})(z) ~.
\ee
Applying equation \erf{eq:cas-loop} to this Casimir
field and substituting \erf{eq:y-def}
yields the quadratic loop equation for the
quiver models
\be
  \zmm^{\rm reg}\Big[~
  \sum_{i,j=1}^r A_{ij} ~\om_i(x) \,\om_j(x)
  ~-~ \frac{2}{N g_s}\sum_{i=1}^r W'_i(x) \, \om_i(x) ~+~ P(x)~\Big] = 0
\labl{eq:ADE-quad-loop}
Here $A_{ij} = (\alpha_i,\alpha_j)$ is the Cartan matrix
of $\lieg$ and we used the identity
$(\alpha_i,\Wh(x)) = W_i(x)$, which follows from the definition
of $\Wh(x)$ below \erf{eq:Wi-poly}. The summand
$(\Wh'(x),\Wh'(x)) / (N g_s)^2$ has been combined with
$2P_2(x)$ to the polynomial $P(x)$, which has maximal degree $2D$,
but at most $D$ coefficients which are not directly
determined by $\Wh(x)$, see section \ref{sec:undet}.
Quadratic loop equations have been
found for similar multi-matrix models in
\cite{Kostov:1995xw, Kostov:1999xi}, and in \cite{Casero:2003gf}
the large $N$ limit of \erf{eq:ADE-quad-loop} was given 
for the $A_r$-case.

\section{Examples}\label{sec:examples}

\subsection{$A_r$--quiver model loop equations at finite $N$}
\label{sec:Ar-ex}

In this section we want to apply the techniques introduced
above to find a set of $r$ loop equations for the 
$A_r$-quiver matrix model. We have
\be
  {\rm exponents~of~} A_r = \{
  1,2,\dots,r{-}1,r\}
\ee
and thus expect the loop equations to have orders
$2,3,\dots, r,r{+}1$.
We will start their construction by introducing a convenient 
set of $r{+}1$ vectors $\{\eps_1,\dots,\eps_{r+1}\}$ 
in $\lieh^*$, corresponding to the weights of the vector 
representation of $A_r$,
\be
  \eps_i = \sum_{k=i}^r \alpha_k - 
  \sum_{k=1}^r \frac{k}{r{+}1} \, \alpha_k 
  \qquad , \quad {\rm for~} i=1,\dots,r{+}1 \quad,
\labl{eq:eps-def}
where $\alpha_1, \dots, \alpha_r$ are the $r$ simple roots
of $A_r$. Using also
$(\alpha_i,\alpha_j) = A_{ij} = 2 \delta_{i,j} - 
\delta_{i,j+1} - \delta_{i,j-1}$ one can verify the
properties
\be
  \alpha_i = \eps_i - \eps_{i+1}
  \quad , \qquad
  \sum_{k=1}^{r+1} \eps_k = 0
  \quad , \qquad
   (\eps_i,\eps_j) = \delta_{i,j} - \frac{1}{r{+}1}  ~.
\labl{eq:eps-prop}

As a further ingredient we will need the commutation
relations of the modes $J^u_k$ with the modes of the
currents $V_{\alpha_k}(z)$, which we will denote by
$V_{\alpha_k,m}$. These follow from \erf{eq:r-fb-block} to be
\be
  [J_m^u, V_{\alpha_k,n}] = (u,\alpha_k) V_{\alpha_k,m{+}n}
  \quad , ~~ {\rm where} ~~
  V_{\alpha_k,m} = 
  \oint_{\gamma_0} \frac{dz}{2\pi i} \, z^m  \, V_{\alpha_k}(z) ~.
\labl{eq:JV-com}
In writing the contour integral for the modes
$V_{\alpha_k,m}$ we used that $V_{\alpha_k}(z)$ has 
conformal weight one. Using \erf{eq:eps-prop} and
\erf{eq:JV-com}, a short calculation shows that, 
for any $\mu\in\cbb$,
\be
  \big[V_{\alpha_k,0} , (\mu+J^{\eps_k}_{-1})(\mu+J^{\eps_{k+1}}_{-1}) \big]
  = J^{\alpha_k}_{-1} V_{\alpha_k,-1} - V_{\alpha_k,-2} ~.
\labl{eq:V-mu-J-mu-J}
Proceeding analogously as in 
\cite{Fateev:1987zh, Bouwknegt:1992wg} 
we define a state
$|R_{r+1}\rangle$ (closely connected to a quantum Miura
transformation),
\be
  |R_{r+1}\rangle = \prod_{i=1}^{r+1} (\mu + J^{\eps_i}_{-1}) 
  |0\rangle ~,
\labl{eq:R-state}
which will serve as a generating function for $r$ Casimir fields.

In fact, consider a field $W(z)$ of the form \erf{eq:W-by-J},
but without derivatives.
Writing out the definition of the normal ordered products, one
verifies
\be
  W(0)|0\rangle = \sum_{a_1,\dots,a_s}^r d_{a_1\dots a_s}
  J^{(a_1)}_{-1}\cdots J^{(a_s)}_{-1} |0\rangle ~.
\ee
In the same way, the field $R_{r+1}(z)$ corresponding to
\erf{eq:R-state}, i.e.\ the field with the property
$R_{r+1}(0)|0\rangle = |R_{r+1}\rangle$ is just given by
the normal ordered product
\be
  R_{r+1}(z) = (A^1 A^2 \dots A^{r+1})(z)
  \qquad {\rm where} \quad
  A^k(z) = \mu \one + J^{\eps_k}(z) ~.
\ee

We would like to show that, for any $\mu\in\cbb$, 
$R_{r+1}(z)$ is a Casimir field. To this end we
verify definition 1 given in section \ref{sec:higher}.
Consider the transformations
\bea \displaystyle
  V_{\alpha_k,0} |R_{r+1}\rangle
  \\[5pt] \displaystyle
  = \prod_{i=1}^{k-1} (\mu + J^{\eps_i}_{-1})
  \big[V_{\alpha_k,0} , (\mu+J^{\eps_k}_{-1})(\mu+J^{\eps_{k+1}}_{-1}) \big]
  \prod_{i=k+2}^{r+1} (\mu + J^{\eps_i}_{-1}) |0\rangle
  \\[5pt] \displaystyle
  =\prod_{i=1}^{k-1} (\mu + J^{\eps_i}_{-1})
  (J^{\alpha_k}_{-1} V_{\alpha_k,-1} - V_{\alpha_k,-2})
  \prod_{i=k+2}^{r+1} (\mu + J^{\eps_i}_{-1}) |0\rangle
  \\[5pt] \displaystyle
  = \underset{i\notin\{k,k{+}1\}}{\prod_{i=1}^{r+1}}
  (\mu + J^{\eps_i}_{-1}) 
  (J^{\alpha_k}_{-1} V_{\alpha_k,-1} - V_{\alpha_k,-2})
  |0\rangle ~.
\eear\labl{eq:V0R}
Here, in the first step we used that by
\erf{eq:JV-com} $V_{\alpha_k,0}$ has nontrivial 
commutation relations only with $J^{\eps_k}_{-1}$
and $J^{\eps_{k+1}}_{-1}$. In the second step \erf{eq:V-mu-J-mu-J}
is substituted, and finally this part of the expression
is commuted past the right product, using that
$V_{\alpha_k,m}$ has trivial 
commutation relations with $J^{\eps_i}_{-1}$ if
$i\ge k{+}2$.

Next we will investigate the states
$J^{\alpha_k}_{-1} V_{\alpha_k,-1}|0\rangle$ and
$V_{\alpha_k,-2}|0\rangle$, and show that they are
in fact equal. This will imply that \erf{eq:V0R}
is equal to zero, and hence 
$R_{r+1}(z)$ is indeed a Casimir field.

First note that by definition
\be
  |\alpha_k\rangle = V_{\alpha_k}(0) |0\rangle
  = \oint_{\gamma_0} \frac{dz}{2\pi i} \,\frac{1}{z} \, 
  V_{\alpha_k}(z) |0\rangle = V_{\alpha_k,-1} |0\rangle ~.
\ee
Further, the state $V_{\alpha_k,-2}|0\rangle$ has $L_0$-eigenvalue
$2$ and charge $\alpha_k$. However, all states with
$L_0$-eigenvalue $2$ and charge $\alpha_k$ are of the form
$J_{-1}^u |\alpha_k\rangle$, for $u\in\lieh^*$. To find which
is the correct value of $u$, we compose both states with
$\langle \alpha_k|J_1^v$ from the left.
Using \erf{eq:r-fb-block} and \erf{eq:JV-com} one easily checks
\be
  \langle \alpha_k|J_1^v J_{-1}^u |\alpha_k\rangle
  = (v,u) \langle \alpha_k |\alpha_k\rangle
  \quad {\rm and} \quad
  \langle \alpha_k|J_1^v V_{\alpha_k,-2}|0\rangle
  = (v,\alpha_k) \langle \alpha_k |\alpha_k\rangle ~.
\ee
For this to hold true for any $v\in\lieh^*$ one needs to have
$u=\alpha_k$. We have thus established
\be
  V_{\alpha_k,-2}|0\rangle = J_{-1}^{\alpha_k} |\alpha_k\rangle ~.
\labl{eq:V2=JV1}

Knowing that $R_{r+1}(z)$ is a Casimir field for 
any $\mu \in \cbb$, we can expand in $\mu$ and find the
individual Casimir fields $W^{(s)}(z)$ of spin
$s = 1,\dots,r{+}1$ as, see 
\cite{Fateev:1987zh} and also 
e.g.\ \cite{Fukuma:1990yk, Kharchev:1992iv, Bouwknegt:1992wg},
\be
  W^{(s)}(z) = 
  \sum_{1\le i_1 < \dots < i_s \le r+1}
  (J^{\eps_{i_1}} \cdots J^{\eps_{i_s}})(z) ~.
\ee
Note that $W^{(1)}=0$ since the $\eps_k$ sum to zero.
We can now apply the relation \erf{eq:cas-loop}
between Casimir fields and loop equations to
all of the $W^{(s)}(z)$. The result can again be
conveniently written in terms of a generating
function
\be
  \zmm^{\rm reg}\!\Big[~~
  \prod_{k=1}^{r+1}\!\big(\mu{+}y_{\eps_k}\!(x)\big) ~+~ 
  \sum_{k=0}^{r+1} \mu^{r+1-k} P_k(x)~~\Big] ~=~ 0 ~~,
\labl{eq:Ar-quiv-loop}
where $P_k(x)$ is a polynomial of degree $\le k D$.
The functions $y_{\eps_k}(x)$ are expressed in terms
of the resolvents and the potential according to
\erf{eq:y-def}. Setting $\om_0(x) = 0 = \om_{r+1}(x)$ 
we can write
\be
  y_{\eps_k}(x) = \om_k(x) - \om_{k-1}(x) + 
  t_k(x)~.
\labl{eq:ydef}
Using \erf{eq:eps-def} together with 
$(\alpha_i,\Wh'(x)) = W'_i(x)$ we find that the tree level 
potentials of quiver model action \erf{eq:quiver-action}
enter in the combination
\be
  t_k(x) = -\frac{1}{g_s N} (\eps_k,\Wh'(x)) =  \frac{1}{g_s N}\Big(
    \sum_{i=1}^r \frac{i}{r{+}1} W'_i(x) - \sum_{i=k}^r W'_i(x)\Big)~.
\labl{eq:tdef}
Note that \erf{eq:ydef} can easily be inverted,
\be
  \om_i(x) = \sum_{k=1}^i \Big( y_{\eps_k}(x)
  -t_k(x) \Big) ~.
\ee
Expanding \erf{eq:Ar-quiv-loop} in $\mu$ gives rise
to $r$ loop equations for the $A_r$-quiver matrix model, 
which hold at finite $N$. The equations arising at the powers
$\mu^{r+1}$ and $\mu^r$ are trivial.

\subsection{A closer look at the cubic and quartic loop equations}

The cubic loop equation of the $A_r$ model corresponds to the coefficient of
$\mu^{r-2}$ in the $\mu$-expansion of \erf{eq:Ar-quiv-loop}. 
A priori it is given as an ordered sum over 
$y_{\eps_i} y_{\eps_j} y_{\eps_k}$. The ordered sum can however
be simplified
by rewriting it in terms of sums over the full index range and
using $\sum_{i=1}^{r+1} y_{\eps_i} = 0$. Explicitly
\be
\sum_{1 \le i < j < k \le r+1} y_{\eps_i} y_{\varepsilon_j} 
y_{\varepsilon_k}  = \frac{1}{3!} \sum_{i,j,k=1}^{r+1} y_{\varepsilon_i} 
y_{\varepsilon_j} y_{\varepsilon_k} - \frac{1}{2} \sum_{i,k=1}^{r+1} 
(y_{\varepsilon_i})^2 y_{\varepsilon_k} + \frac{1}{3}
\sum_{i=1}^{r+1} (y_{\varepsilon_i})^3 \,,
\labl{eq:Ar-cubic-ordered}
where on the rhs all but the last term sum to zero. This yields
the following form for the cubic loop equation, valid for general
$r$,
\be
\zmm^{\rm reg}\Big[\sum_{i=1}^{r+1} (y_{\varepsilon_i})^3 + 3P_3\Big] = 0 \,.
\labl{eq:Ar-cubic}
Then, using $\omega_0 = 0 = \omega_{r+1}$, $\varepsilon_i = 
\alpha_i + \varepsilon_{i+1}$ and defining $\eta = N g_s$, one obtains
\bea\displaystyle
\zmm^{\rm reg}\Big[\sum_{i=1}^{r} \Big( \omega_{i-1} \omega_i^2 - 
\omega_{i-1}^2 \omega_i + \frac{1}{\eta} W_i' \big( \omega_i^2 -
\frac{1}{\eta} \omega_i W_i' \big) 
\\[5pt] \displaystyle
     \hspace{5em}
+ \frac{2}{\eta}
\big( \omega_i^2 - \omega_i \omega_{i+1} -  \frac{1}{\eta}
\omega_i W_i' \big) (\varepsilon_{i+1}, \Wh') \Big) + Q_3\Big] = 0 \,,
\eear\labl{eq:A3-cub-1}
for some polynomial $Q_3(x)$ of degree $\leq 3 D$.
This form of the cubic can be transformed further
by making use of the quadratic loop equation \erf{eq:ADE-quad-loop}.
Substituting $\eps_{i+1} = \eps_2 - \sum_{k=2}^i \alpha_k$ into
\erf{eq:A3-cub-1}, the coefficient of $(\varepsilon_{2}, \Wh')$ is
just a polynomial (due to the quadratic loop equation) and can
be absorbed into $Q_3$, resulting in
\bea\displaystyle
 \zmm^{\rm reg}\Big[\sum_{i=1}^{r} \Big( \omega_{i-1} \omega_i^2 - 
\omega_{i-1}^2 \omega_i + \frac{1}{\eta} W_i' \big( \omega_i^2 -
\frac{1}{\eta} W_i' \omega_i \big) \Big)
\\[5pt]\displaystyle \hspace{5em}
- \frac{2}{\eta} \sum_{i=2}^r \big( \omega_i^2 - \omega_i \omega_{i+1} -  
\frac{1}{\eta}  W_i' \omega_i\big) \sum_{k=2}^i W_k' + \tilde Q_3 \Big] = 0 \,,
\eear\ee
where $\tilde Q_3(x)$ is some polynomial of degree $\leq 3 D$.
For $r=2$ this equation reduces to the one in \cite{Klemm:2003cy}.

The quartic loop equation can be treated in the same way, taking the
coefficient of $\mu^{r-3}$ in \erf{eq:Ar-quiv-loop}. For $A_3$ this coefficient
is just $y_{\eps_1} y_{\eps_2} y_{\eps_3} y_{\eps_4}$ and the loop
equation reads explicitly
\bea\displaystyle
 \zmm^{\rm reg}\big[~(3W_1' + 2 W_2' + W_3' - 4 \eta \omega_1 ) 
   (W_1' - 2 W_2' - W_3' - 4 \eta \omega_1 +  4 \eta \omega_2)
\\[2pt]\displaystyle
   \hphantom{ \zmm^{\rm reg}\big[~ }
(W_1' + 2 W_2' +3 W_3' - 4 \eta \omega_3)
(W_1' + 2 W_2' - W_3' - 4 \eta \omega_2 +  4 \eta \omega_3)
\\[2pt]\displaystyle
      \hphantom{ \zmm^{\rm reg}\big[~ }
 -256 \eta^4 P_4~\big] = 0 \,.
\eear\ee
For general $r$ it is convenient to rewrite the ordered sum over
four indices similarly as was done above for three indices, giving
the $A_r$ quartic loop equation in the form
\be
\zmm^{\rm reg}\Big[~\sum_{i,k=1}^{r+1} (y_{\eps_i})^2 
(y_{\eps_k})^2 -2\sum_{i=1}^{r+1} (y_{\eps_i})^4 + 8 P_4~\Big] = 0 \,.
\labl{eq:Ar-quartic}

\subsection{$A_r$--quiver model loop equations at large $N$}
\label{sec:Ar-large-N}

In the large $N$ limit (keeping $N_i/N$ and $g_s N$ fixed)
the $n$-point functions in the matrix model factorise into
products of $n$ one-point functions. The loop equations
encoded in \erf{eq:Ar-quiv-loop} become algebraic.
Abbreviating 
$\tilde y_{\eps_k}(x) = \zmm^{\rm reg}[y_{\eps_k}(x)] / \zmm^{\rm reg}$
and replacing $\mu\rightarrow z$ we can write
\be
  \prod_{k=1}^{r+1} \!\big(\,z{+}\tilde y_{\eps_k}(x)\,\big)
  ~=~ -\sum_{k=0}^{r+1} P_k(x) z^{r+1-k} ~,
\labl{eq:Ar-largeN}
where again $P_k(x)$ is a polynomial of degree $\le k D$.
It follows that, given the polynomials
$P_k(x)$, the $\tilde y_{\eps_k}(x)$ are minus the $r{+}1$ roots of
the polynomial in $z$ on the rhs of \erf{eq:Ar-largeN},
i.e.\ minus the $r{+}1$ solution to the equation
\be
  z^{r+1} - \sum_{k=0}^{r-1} P_{r+1-k}(x) z^k = 0 ~,
\ee
where we also used that \erf{eq:Ar-largeN} forces
$P_0(x)=-1$ and $P_1(x)=0$.
This also proves a claim in \cite{Dijkgraaf:2002vw}
that the loop equations of the $A_r$-quiver matrix
models reproduce, in the large $N$ limit, a deformed
reduction to one complex dimension of the singular, 
non-compact Calabi-Yau threefold geometry associated to 
the $A_r$-quiver. Explicitly, the non-compact 
Calabi-Yau three-fold is given by the fibration of 
an $A_r$-singularity over the complex $x$-plane,
\cite{KM92, Cachazo:2001gh, Dijkgraaf:2002vw},
\be
  u^2 + v^2 + \prod_{k=1}^{r+1} (z+t_k(x)) = 0 ~~.
\label{eq:arcygeom}
\ee
Here the $t_k(x)$ as given by \erf{eq:tdef} are 
polynomials, fixed by the tree level potentials, 
and hence the geometry reduced with respect to the $u$ and 
$v$ direction is a nodal curve. In the quantum geometry 
as described by the matrix model loop equation the 
$t_k(x)$ are replaced by $\tilde y_{\epsilon_k}(x)$ and 
the $DR_+$ parameters dicussed in section \ref{sec:undet} 
are turned on. 
The curve does not factorise algebraically any more 
and the double points get resolved. Note that the number
of parameters $DR_+$, which are not fixed at tree level 
is sufficient to resolve all double points. 

Topological string amplitudes of the B-model and exact gauge 
theory quantities are calculated in terms of periods of the 
resolved curve with respect to the meromorphic differentials, which are reductions 
of the Calabi-Yau $(3,0)$ form over $r$ cycles of \erf{eq:arcygeom}. 
They are given by 
\be
  \eta_i ~=~ \tilde y_{\alpha_i}(x){\rm d} x ~=~
  \Big(\sum_{j=1}^r A_{ij}\,\om_j(x) - \frac{1}{N g_s}
  W'_i(x) \Big){\rm d} x ~,
\labl{eq:merodiff}
where we set $u=\alpha_i$ in \erf{eq:y-def} and used
$(\alpha_i,\Wh'(x))=W'_i(x)$.

\medskip 

Note that expanding out \erf{eq:Ar-largeN} gives the
$r$ loop equations as ordered sums,
\be
  \sum_{1 \le i_1 < \dots < i_s \le r+1}  
  \tilde y_{\eps_{i_1}}(x) \dots \tilde y_{\eps_{i_s}}(x) + P_s(x) = 0~~.
\labl{eq:Ar-largeN-ordered}
A corresponding equation has been conjectured in \cite{Casero:2003gf}
to arise from an analysis of the Konishi anomaly in quiver
gauge theories.
Using $\sum_{i=1}^{r+1} \tilde y_{\eps_i}(x) = 0$, equation
\erf{eq:Ar-largeN-ordered} can be rewritten in the more concise form
\be
  \sum_{i=1}^{r+1} \tilde y_{\eps_i}(x)^s ~+~ Q_s(x) ~=~ 0 ~~,
\labl{eq:Ar-largeN-simple}
for some polynomial $Q_s(x)$ of degree $\le s D$. The equivalence of
\erf{eq:Ar-largeN-ordered} and \erf{eq:Ar-largeN-simple} can be seen recursively. One first checks
the statement for $s{=}2$. 
For general $s$ one rewrites the ordered sum in terms of several sums 
over the full index range and uses the already established
identities \erf{eq:Ar-largeN-simple} of degree less than $s$.
For example in the case $s{=}3$ one rewrites the ordered
sum as in \erf{eq:Ar-cubic-ordered}, 
where only the last term survives. In the
case $s{=}4$ 
the same procedure leads to the large $N$ form of \erf{eq:Ar-quartic};
using the case $s{=}2$, one can replace
$\sum_{i,j=1}^{r+1} \tilde y_{\eps_i}(x)^2 
\tilde y_{\eps_j}(x)^2 = Q_2(x)^2$ 
to obtain the form \erf{eq:Ar-largeN-simple}.

\subsection{$D_r$--quiver model loop equations at finite $N$}

Next we turn to the investigation of $D$-series. For the Lie
algebra $D_r$ with $r\ge 3$ we have
\be
  {\rm exponents~of~} D_r ~=~ \{\, r{-}1 \,,\, 1 , 3 , 
  \dots,2r{-}3 \,\}  ~~.
\ee
Accordingly we expect a generating set of Casimir fields 
of spin $r$ and spins $2,4,\dots,2r{-}2$. The procedure of 
\cite{Lukyanov:gg}, which we will recall below, is to start 
by constructing the Casimir field $W^{(r)}(z)$ of spin $r$ and then find
the fields of spin $2, 4, \dots 2r{-}2$ as elements of
the OPE $W^{(r)}(z) W^{(r)}(w)$. 

As for $A_r$ we first need to choose a convenient set
of vectors $\eps_i$ in $\lieh^*$. In the case of $D_r$ 
we choose $r$ such vectors, which form a basis of $\lieh^*$.
Let the simple roots of $D_r$ be numbered such that the
Cartan matrix $A_{ij} = (\alpha_i,\alpha_j)$ is of the form
\be
  A_{ij} = 2\delta_{i,j} - \delta_{i,j+1} - \delta_{i,j-1}
  + \delta_{i,r} ( \delta_{j,r-1} {-} \delta_{j,r-2} )
  + \delta_{j,r} ( \delta_{i,r-1} {-} \delta_{i,r-2} ) ~.
\ee
The vectors $\eps_i$, for $i=1,\dots,r$ are defined as
\be
  \eps_i ~=~ \sum_{k=i}^{r-2} \alpha_k ~+~ 
  \tfrac12 \big(\alpha_{r} + (-1)^{\delta_{i,r}} \alpha_{r-1} \big) ~~,
\ee
where for negative range the sum is taken to be zero.
The vectors $\eps_1,\dots,\eps_r$ have the properties
\be
  (\eps_i,\eps_j) = \delta_{i,j} ~,~~
  \alpha_i = \eps_i - (1{-}\delta_{i,r})\eps_{i+1} + 
    \delta_{i,r} \eps_{r-1} ~.
\labl{eq:D-eps-prop}
Using the relations \erf{eq:JV-com} and 
\erf{eq:V2=JV1} (which are not specific to $A_r$ but valid
for any simply laced Lie algebra) one can check the equations,
for $i\le r{-}1$,
\bea\displaystyle
  V_{\alpha_i,0} J^{\eps_i}_{-1} J^{\eps_{i+1}}_{-1} |0\rangle
  = J^{\alpha_i}_{-1} V_{\alpha_i,-1} |0\rangle
    -  V_{\alpha_i,-2} |0\rangle = 0 ~~, 
  \\[5pt]\displaystyle
  V_{\alpha_r,0} J^{\eps_{r-1}}_{-1} J^{\eps_{r}}_{-1} |0\rangle
  = -J^{\alpha_r}_{-1} V_{\alpha_r,-1} |0\rangle
    +  V_{\alpha_r,-2} |0\rangle = 0 ~~.
\eear\ee
Define the field $W^{(r)}(z)$ via
\be
  W^{(r)}(0) |0\rangle = 
  J^{\eps_1}_{-1} J^{\eps_2}_{-1} \cdots J^{\eps_r}_{-1} |0\rangle  ~.
\labl{eq:D-Wr-def}
Since all of the $r$ $J$-modes entering \erf{eq:D-Wr-def} commute
and also each of the screening charge zero modes $V_{\alpha_i,0}$ commutes
with all but two of the $J^{\eps_k}$ one finds
\bea\displaystyle
  V_{\alpha_i,0} W^{(r)}(0) |0\rangle
  = \Big(\underset{k\notin\{ i,i+1\} }{\prod_{k=1}^r} J^{\eps_k}_{-1}\Big)
  V_{\alpha_i,0} J^{\eps_i}_{-1} J^{\eps_{i+1}}_{-1} |0\rangle
  = 0 ~~, \\[5pt]\displaystyle
  V_{\alpha_r,0} W^{(r)}(0) |0\rangle
  = \Big(\prod_{k=1}^{r-2} J^{\eps_k}_{-1} \Big)
  V_{\alpha_r,0} J^{\eps_{r-1}}_{-1} J^{\eps_r}_{-1} |0\rangle
  = 0 ~~.
\eear\ee
Thus $W^{(r)}(z)$ is indeed a Casimir field. Since by translation
invariance also $[V_{\alpha_i,0}, W^{(r)}(z)]=0$, we have in 
particular that, for $i=1,\dots,r$ and all values of $z$,
\be
  V_{\alpha_i,0} W^{(r)}(z) W^{(r)}(0) |0\rangle = 0 ~~.
\ee
It follows that upon writing out the OPE
\be
  W^{(r)}(z) \, W^{(r)}(0) \, |0\rangle ~= \sum_{m=-\infty}^{2r}
  z^{-m} ~ \widetilde W^{(2r-m)}(0) |0\rangle ~,
\labl{eq:tW-OPE}
every coefficient $\widetilde W^{(2r-m)}(0)|0\rangle$ is
a state corresponding to a Casimir field.
This gives an a priori infinite number of Casimir fields, but they will
not all be independent. 
However, one can show that the $W^{(r)}(z)$ together with
$\widetilde W^{(s)}(z)$, $s=2,4,\dots,2r{-}2$ do indeed form
an independent set of Casimir fields, see
\cite{Lukyanov:gg} and e.g.\ \cite{Bouwknegt:1992wg} section 6.3.3.

Let us investigate the fields 
$\widetilde W^{(s)}(z)$, $s=2,4,\dots,2r{-}2$ in more detail.
In terms of the modes $W_m$ of $W^{(r)}(z)$, defined via
$W^{(r)}(z) = \sum_m z^{-m-r} W_m$, the OPE \erf{eq:tW-OPE} 
takes the form
\be
  W^{(r)}(z) W^{(r)}(0) |0\rangle = \sum_{m\in\zbb}
  z^{-m-r} W_m W_{-r} | 0 \rangle ~,
\ee 
so that $\widetilde W^{(s)}(0)|0\rangle = W_{r-s} W_{-r} | 0 \rangle$.
Using the definition of the normal ordered product, it is
not difficult to show that
\bea\displaystyle
  W^{(r)}(z) ~=~ (J^{\eps_1} \cdots J^{\eps_r})(z) ~=\hspace{-1em} 
  \sum_{m_1,\dots,m_r \in \zbb}\hspace{-1em} 
  z^{-(m_1 + \cdots + m_r + r)} ~ \nop J^{\eps_1}_{m_1} \cdots
  J^{\eps_r}_{m_r} \nop ~~, \\[5pt]\displaystyle
  W_m ~=\hspace{-1em}  \underset{m_1+\cdots+m_r=m}{\sum_{m_1,\dots,m_r \in \zbb}}
  \nop J^{\eps_1}_{m_1} \cdots J^{\eps_r}_{m_r} \nop ~~.
\eear\labl{eq:Wm-modes}
where $\nop \cdots \nop$ denotes the usual normal ordering
of modes, such that the positive modes are to the right.
Because of \erf{eq:r-fb-block} and \erf{eq:D-eps-prop} all
the $J$-modes in \erf{eq:Wm-modes} commute and the normal
ordering can be omitted.
If a term in the sum \erf{eq:Wm-modes} for $W_m$ is to
contribute to the product $W_m W_{-r} |0\rangle$, we need
the $m_k$ to be in the set
$S = \{1\} \cup \zbb_{\le -1}$. This gives explicitly
\bea\displaystyle
 \widetilde W^{(s)}(0) |0\rangle
 ~= \hspace{-1em}\underset{m_1+\cdots+m_r = r-s}{\sum_{m_1,\dots,m_r \in S}}
 J^{\eps_1}_{m_1} J^{\eps_1}_{-1} \cdots
 J^{\eps_r}_{m_r} J^{\eps_r}_{-1} |0\rangle 
 \\[5pt]\displaystyle
 =~ \sum_{n=1}^{s/2} ~ \sum_{1 \le i_1 < \dots < i_n\le r}
 \underset{m_1+\cdots+m_n = s-2n}{\sum_{m_1,\dots,m_n \ge 0}}\hspace{-1em}
  J^{\eps_{i_1}}_{-m_1-1} J^{\eps_{i_1}}_{-1} \cdots
 J^{\eps_{i_n}}_{-m_n-1} J^{\eps_{i_n}}_{-1} |0\rangle
\eear\labl{eq:tW-modes}
The second expression for $\widetilde W^{(s)}(0) |0\rangle$
is obtained form the first by cancelling all 
$J_{m_k}^{\eps_k} J_{-1}^{\eps_k}$ where $m_k=1$ and redefining
the other $m_k$ as $m_k \leadsto -m_k{-}1$.

To proceed we will express the higher $J$-modes \erf{eq:tW-modes}
in terms of derivatives. Noting that 
$(L_{-1})^n J_{-1} |0\rangle = n! J_{-n-1} |0\rangle$ we see that
the field corresponding to the state $J_{-n-1} |0\rangle$ is
$\tfrac{1}{n!} \partial^n J(z)$. Using this, we finally find
\be
 \widetilde W^{(s)}(z)
 = \sum_{n=1}^{s/2} ~ \sum_{1 \le i_1<\dots<i_n\le r}
 \underset{m_1+\cdots+m_n = s-2n}{\sum_{m_1,\dots,m_n \ge 0}}\hspace{-1em}
 \frac{
 (J^{\eps_{i_1}} \partial^{m_1} J^{\eps_{i_1}} \cdots
 J^{\eps_{i_n}} \partial^{m_n} J^{\eps_{i_n}})(z)}{m_1! \cdots m_n!}~.
\ee
With the help of \erf{eq:cas-loop},
we can now write down the $r$ loop equations resulting from
the Casimir fields 
$\{ W^{(r)}(z) , \widetilde W^{(2)}(z), \widetilde W^{(4)}(z),
\dots , \widetilde W^{(2r-2)}(z) \}$. For $W^{(r)}(z)$ one finds
\be
  \zmm^{\rm reg}\big[~ y_{\eps_1}(x) y_{\eps_2}(x) \cdots y_{\eps_r}(x)
  + P_r(x)~\big] = 0~~,
\labl{eq:D-r-loop}
where $P_r(x)$ is a polynomial of degree $\le r D$, while for
$\widetilde W^{(s)}(z)$ one gets
\be
  \zmm^{\rm reg}\Big[~
  \sum_{n=1}^{s/2} N^{2n-s} \hspace{-1.5em}
  \sum_{1 \le i_1<\dots<i_n\le r}
  \underset{m_1+\cdots+m_n = s-2n}{\sum_{m_1,\dots,m_n \ge 0}}
  \prod_{k=1}^n \frac{y_{\eps_{i_k}}(x) \, \partial^{m_k}
  y_{\eps_{i_k}}(x)}{m_k!} ~+~ \tilde P_s(x) ~\Big] = 0 ~~,
\labl{eq:D-loop}
where $\tilde P_s(x)$ is a polynomial of degree $\le s D$.
The $y_{\eps_k}(x)$ are related to the resolvents via
\erf{eq:y-def}. Explicitly, defining also $\om_0(x)\equiv 0$,
\bea\displaystyle
  y_{\eps_k}(x) = \om_k(x) - \om_{k-1}(x) + \delta_{k,r-1} \om_r(x)
  +t_k(x) ~, \\[5pt]\displaystyle
  t_k(x)=  -\frac{1}{N g_s} (\eps_k,W'(x)) =-\frac{1}{N g_s} \Big(
  \sum_{i=k}^{r-2} W'_i(x) + \tfrac12\big( W'_{r}(x) + 
  (-1)^{\delta_{k,r}}W'_{r-1}(x) \big) \Big)~.
\eear 
\labl{eq:tdefdr}
Again this relation can be inverted to give the resolvents
$\om_i$ in terms of the $y_{\eps_i}$,
\be
  \om_i(x) ~=~ \frac12\big(2{-}\delta_{i,r-1}{-}\delta_{i,r}\big)
  \sum_{k=1}^{i} \Big( y_{\eps_k}(x) -t_k(x)  \Big)
  ~-~ \frac12 \, \delta_{i,r-1} \Big( y_{\eps_r}(x) - t_k(x)\Big) ~~.
\ee
Note that in contrast to $A_r$, the $D_r$--loop equations 
\erf{eq:D-loop} in general 
contain derivatives of the variables $y_{\eps_k}(x)$.
These derivatives appear because the corresponding Casimir
field \erf{eq:tW-modes} contains modes $J_{-m}$ with $m>1$.
Recall that this was not the case in the generating function
\erf{eq:R-state} for the $A_r$ Casimir fields.

However, three of the Casimir fields can be expressed through
the modes $J^{\eps_k}_{-1}$ alone. The first is, by definition,
$W^{(r)}(z)$ in \erf{eq:D-Wr-def}. The second and the third
are the stress tensor $T(z)$ and a spin 4 field $U(z)$, 
defined as
\bea\displaystyle
  T(0)|0\rangle = \tfrac12 \sum_{i=1}^r J^{\eps_i}_{-1}
  J^{\eps_i}_{-1} |0\rangle = 
  \tfrac12 \widetilde W^{(2)}(0)|0\rangle ~~,
  \\[5pt]\displaystyle
  U(0)|0\rangle = 
  \tfrac12 \sum_{1\le i<j \le r} J^{\eps_i}_{-1}J^{\eps_i}_{-1}
  J^{\eps_j}_{-1}J^{\eps_j}_{-1} |0\rangle
  - \tfrac14 \sum_{i=1}^r J^{\eps_i}_{-1}J^{\eps_i}_{-1}
  J^{\eps_i}_{-1}J^{\eps_i}_{-1}|0\rangle  
  \\[5pt]\displaystyle
  \hspace{8em} = 
  \widetilde W^{(4)}(0)|0\rangle - L_{-2} L_{-2} |0\rangle ~.
\eear\labl{eq:D-TU-def}
The last equality can be verified using
\be
  L_m = \frac12 \sum_{k\in\zbb} \sum_{i=1}^r \nop 
  J^{\eps_i}_{k} J^{\eps_i}_{m-k}\nop ~~.
\ee
Further, 
to see that $U(z)$ is indeed a Casimir field one can use
that $\widetilde W^{(4)}(z)$ is a Casimir field and that
the screening charge zero modes commute with the Virasoro
generators, $[V_{\alpha_i,0} , L_m] = 0$. This in turn follows
from the commutation relation of the $L_m$ with the modes
$K_n$ of any primary spin
one field $K(z)$, which read $[L_m , K_n ] = -n K_{n+m}$. 

The Casimir fields $W^{(r)}(z)$, $T(z)$ and $U(z)$ thus
result in loop equations which do not involve derivatives.
Note that this fits well with the observation that
$D_3 = A_3$, as Lie algebras, and the three Casimir fields
of $A_3$ with spins 2, 3 and 4 do not contain derivatives.
For the $\widetilde W^{(s)}(z)$ with $s\ge 6$, linear
combinations similar as in \erf{eq:D-TU-def} do not seem
to exist.

\subsection{$D_r$--quiver model loop equations at large $N$}
\label{sec:Dr-large-N}

When taking the large $N$ limit of \erf{eq:D-loop} only
the coefficient of $N^0$ survives. For this coefficient
we have $n=s/2$, so that it does not contain any derivatives.
Abbreviating again
$\tilde y_{\eps_k}(x) = \zmm^{\rm reg}[y_{\eps_k}(x)] / \zmm^{\rm reg}$
the large $N$ limit of the loop equations 
\erf{eq:D-r-loop} and \erf{eq:D-loop} reads
\be
  \tilde y_{\eps_1}(x) \tilde y_{\eps_2}(x) \cdots \tilde y_{\eps_r}(x)
  ~=~ - P_r(x)
\labl{eq:D-large-N-1}
and
\be
  \sum_{1 \le i_1<\dots<i_{s/2} \le r}
  \hspace{-2em}
  {\tilde y_{\eps_{i_1}}(x)}^2 
  \cdots
  {\tilde y_{\eps_{i_{s/2}}}(x)}^2 
  ~=~ -\tilde P_s(x) ~~,
\labl{eq:D-large-N-2}
where $s=2,4,\dots,2r-2$. As opposed to their finite
$N$ form, the large $N$ loop equations can easily be 
written in terms of a generating function,
\be
  \prod_{k=1}^r (z + {\tilde y_{\eps_k}(x)}^2) 
  = - \sum_{k=0}^r \tilde P_{2r-2k}(x) z^k ~~,
\labl{eq:D-large-N-gen}
where we set $\tilde P_0(x) = -1$ and 
$\tilde P_{2r}(x) = -P_r(x)^2$. 
Note that \erf{eq:D-large-N-gen} is weaker than the
set of equations \erf{eq:D-large-N-1} and \erf{eq:D-large-N-2} 
because the coefficient
of $z^0$ just gives the square of relation  
\erf{eq:D-large-N-1}. From \erf{eq:D-large-N-gen} we see that
the $\tilde y_{\eps_k}(x)$ are equal to $\pm i$ times the 
$r$ zeros of the polynomial in $z$ given by
$\sum_{k=0}^r \tilde P_{2r-2k}(x) z^k = 0$ 
with the constraint that the signs of
the $\tilde y_{\eps_k}(x)$ have to be chosen such
that equation \erf{eq:D-large-N-1} holds.

Similar as in the $A_r$ case, the large $N$ loop
equations of the $D_r$-quiver matrix model is the 
deformation of the reduction of a Calabi-Yau three-fold
to a nodal curve.  The non-compact threefold is 
a fibration of a $D_r$-singularity over the complex 
$x$-plane with defining equation
\cite{KM92, Cachazo:2001gh, Dijkgraaf:2002vw},
\be
  u^2 + v^2 z + 
  \frac{1}{z} \Big(
  \prod_{k=1}^r (z+ t_k(x)^2) 
  -\prod_{k=1}^r t_k(x)^2 \Big) + v \prod_{k=1}^r t_k(x) = 0 ~~,
\ee
where $t_k(x)$ are given by \erf{eq:tdefdr}. The deformation of 
nodes is achieved by by replacing $t_k(x)$ with $\tilde y_{\epsilon_k}(x)$ 
and turning on the $DR_+$ parameters. Note that due to the $Z_2$ 
symmetry of  \erf{eq:D-large-N-gen} pairs of douple points are 
resolved by this. The reduction of the holomorphic $(3,0)$ 
on the curve gives meromorphic forms as in \erf{eq:merodiff}.

\section{Relation to Casimir algebras}\label{sec:casimir}

The aim of this section is to relate the
calculations in sections \ref{sec:ADE-FB} and \ref{sec:examples} to a
special form of $\Wc$-algebras, the so-called 
Casimir algebras. We will start by some general
comments on $\Wc$-algebras.

\subsection{Some generalities on $\Wc$-algebras}

By a chiral algebra one denotes a subsector of a full
conformal field theory which consists only of holomorphic
fields, i.e.\ of fields $\phi(z,\bar z)$ which obey
$\partial/\partial \bar z \, \phi(z,\bar z) = 0$. The chiral algebra
need not contain all such fields, the only requirements are
that it closes under the OPE and that it contains the
conformal stress tensor $T(z)$.

Chiral algebras can be studied as objects on their own
right, without reference to the original full CFT. This
leads to the notion of a conformal vertex algebra, see
e.g.\ \cite{Frenkel} for a mathematical exposition of 
the subject. In fact, a standard approach to study CFTs is
to start with a chiral algebra and to use its representation 
theory to construct the full CFT. This has lead for example
to the Virasoro minimal models and the WZW-models, see 
e.g.\ \cite{DiFrancesco:nk}.

For the Virasoro minimal models, the chiral algebra is
just the Virasoro algebra, generated only by the stress
tensor $T(z)$ itself. For WZW models the chiral algebra
is a current algebra, that is, it is generated by fields
of spin one. An obvious generalisation is
to study chiral algebras that are still generated by
a {\em finite} number of fields, but where the fields
can have any integer spin. These are the $\Wc$-algebras;
an extensive review can be found in \cite{Bouwknegt:1992wg}.
$\Wc$-algebras made their appearance in \cite{Zamolodchikov:wn},
where a chiral algebra generated by $T(z)$ and an 
additional field of spin three was investigated. 

A qualitative distinction between the Virasoro or
current algebras and general $\Wc$-algebras is that
the OPE of two generating fields is no longer
a linear expression in the generators. On the level
of modes this implies that the algebra of modes
of the generating fields is not a Lie algebra.
The commutator of two modes can contain 
(infinite sums of) products of modes.

$\Wc$-algebras appear in the context of matrix
models and integrable hierarchies in the form
of $\Wc$-constraints. In this case one has a
set of differential operators $W_n$ which annihilate
the $\tau$-function of the hierarchy and which
do not form a Lie-algebra. Instead their commutators
can give non-linear combinations of the $W_n$, 
and the resulting structure is that of the
mode algebra of a conformal $\Wc$-algebra 
\cite{Dijkgraaf:1990rs, Fukuma:1990jw}.

A link between the $\Wc$-constraints in the matrix
model context and the $\Wc$-algebras of conformal
field theory is provided by the free boson representation
of the matrix model, see 
\cite{Kharchev:1992iv, Morozov:hh, Kostov:1999xi}.
In this paper we worked in the free boson representation
right from the start.
Another link, this time more generally between 
the Hirota bilinear equations for 
$\tau$-functions and $\Wc$-algebras, is given by
the orbit construction, see e.g.\ \cite{KdV-Thesis}
and references therein.

\subsection{Casimir algebras and WZW-models}

The $\Wc$-algebras we are interested in for the purposes
of this paper are of a special type, called
Casimir algebras. These are defined as follows, see 
\cite{Bais:1987dc} as well as \cite{Bouwknegt:1992wg} and
references therein. 
Consider a WZW-model $\hat\lieg_k$ at level $k$ for a Lie algebra 
$\lieg$. The modes $T^i_m$ of the spin one currents
$T^i(z)$ that generate the chiral algebra span an
affine Lie algebra at level $k$, which we will also denote
by $\hat\lieg_k$,
\be
  [\,T^i_m \,,\, T^j_n \,]
  ~=~ \sum_{\ell=1}^{\dim\,\lieg} f^{ij}_{~~\ell} \, T^\ell_{m+n} +
  k m \, K(T^i,T^j) \, \delta_{m{+}n,0} ~,
\ee
where $T^i$ denote the generators of the underlying
finite Lie algebra $\lieg$, the $f^{ij}_{~~\ell}$ are the
structure constants of $\lieg$ and $K(\cdot,\cdot)$ is the
Killing form on $\lieg$.
Note that $\lieg$ is canonically embedded into $\hat\lieg_k$
via $T^i \mapsto T^i_0$. Denote this embedding by $\iota$.
Similar to the coset construction
of conformal field theory we can consider all elements
in $\hat\lieg_k$ that commute with all elements of
$\iota(\lieg)$.

\medskip\noindent
{\bf Definition:} The {\em Casimir algebra}
$\Wc[\hat\lieg_k / \lieg]$ is defined as the (vertex--) subalgebra
of the chiral algebra $\hat\lieg_k$ obtained by taking all fields
$\phi(z)$ of $\hat\lieg_k$ that obey
$[T^i_0, \phi(z)]=0$ for all generators $T^i$ of $\lieg$.
\medskip

Via the state-field correspondence we can give an equivalent
characterisation of $\Wc[\hat\lieg_k / \lieg]$ in terms of 
states in the vacuum module of $\hat\lieg_k$,
\be
  [T^i_0, \phi(z)] = 0 
  \quad \Leftrightarrow \quad
  T^i_0 \phi(0) |0\rangle = 0~.
\labl{eq:Wgkg-state}
The chiral algebra $\hat\lieg_k$ is spanned as a vector
space by fields of the form
\be
   W^{i_1\dots i_n}_{m_1\dots m_n}(z)
   = (\partial^{m_1} T^{i_1} \cdots \partial^{m_n} T^{i_n})(z) ~.
\ee
One can verify (see \cite{Bouwknegt:1992wg})
that, for generic level $k$, the linear combination
\be
  W_{m_1\dots m_n}(z) = \sum_{i_1,\dots,i_n}
  d_{i_1\dots i_n} W^{i_1\dots i_n}_{m_1\dots m_n}(z)
\ee
is in $\Wc[\hat\lieg_k / \lieg]$ if and only if
$\sum_{i_1,\dots,i_n}
d_{i_1\dots i_n} T^{i_1} \cdots T^{i_n}$ is in the centre
of the universal enveloping algebra of $\lieg$, 
i.e.\ if it is a Casimir element of $\mathsf{U}(\lieg)$.
This is the reason for the terminology `Casimir algebra'.
For certain specific values of $k$, however, one looses the `only if'
in the above relation (just take the free boson stress tensor
in the free boson realisation of $\hat\lieg_1$ as an example).

\subsection{Casimir fields and Casimir algebras}

We would like to find the relation between the notion
of a `Casimir field' used in sections
\ref{sec:ADE-FB} and \ref{sec:examples} and the Casimir algebras introduced
above.

Choose a Cartan-Weyl basis for $\lieg$. In the free field
realisation of $\hat\lieg_1$, the free boson currents
$J^{(a)}(z)$ provide the $\hat\lieg_1$--fields
$H^i(z)$, where $H^i$ denotes an element of the Cartan
subalgebra of $\lieg$. The $\hat\lieg_1$--fields
$E^{\alpha_i}(z)$, for the ladder operators
$E^{\alpha_i} \in \lieg$ and $\alpha_i$ a simple root,
are given by the free boson vertex operators
$V_{\alpha_i}(z)$, up to a cocycle factor, which we do 
not spell out explicitly, see e.g.\ \cite{DiFrancesco:nk}
section~15.6.3.
We would like to establish
\be
  \phi(z) {\rm~is~a~Casimir~field}
  \qquad \Leftrightarrow \qquad
  \phi(z) \in \Wc[\hat\lieg_1 / \lieg] ~.
\ee
It is enough to check the defining condition in \erf{eq:Wgkg-state} for the 
modes $E^{\alpha_i}_0$ and $H^i_0$ for $i=1,\dots,r$. If $\phi(z)$ is a 
Casimir field, then $J^{(a)}_0 \phi(0) |0\rangle = 0$ 
since by construction the state $\phi(0)|0\rangle$ is in the 
vacuum module of the $U(1)^r$--algebra and thus has $U(1)^r$-charge zero. 
Furthermore, by definition~1 in section \ref{sec:higher} 
we have $V_{\alpha_i,0} \phi(0) |0\rangle = 0$, so that indeed
$\phi(z) \in \Wc[\hat\lieg_1 / \lieg]$. 

In the converse direction we have to be more careful, because
in the definition we used, Casimir fields are build {\em only}
from the $\hat\lieg_1$--currents $H^i(z)$, corresponding to the
Cartan subalgebra of $\lieg$, while the fields in $\Wc[\hat\lieg_1 / \lieg]$
are constructed from {\em all} generators $T^i(z)$ in $\hat\lieg_1$.
Suppose $\phi(z) \in \Wc[\hat\lieg_1 / \lieg]$. Then because
$H^i_0 \phi(0)|0\rangle=0$ we know that the state
$\phi(0) |0\rangle$ is indeed in the vacuum module of the 
$U(1)^r$--algebra and can thus be expressed in terms of the
modes $J^{(a)}_{-m}$ alone. The remaining conditions
$E^{\alpha_i}_0 \phi(0)|0\rangle=0$ then imply that $\phi(z)$
is a Casimir field.

\medskip

In the $\lieg = A_r$ and $\lieg = D_r$ examples discussed in 
section \ref{sec:examples} we have constructed $r$ fields of 
$\Wc[\hat\lieg_1 / \lieg]$. One can now wonder if by repeatedly
taking normal ordered products, these fields (and their
derivatives) generate all of
$\Wc[\hat\lieg_1 / \lieg]$. There are arguments
(see \cite{Bouwknegt:1992wg} and references therein), using
the Drinfeld-Sokolov reduction and the character technique,
that this in indeed the case.

\appendix

\section{Appendix: Integral form of $\mathcal{W}$-constraints}\label{sec:app-lim}

In this appendix we establish equation \erf{eq:intW=0}.
Denote by $I(L)$ the integral in \erf{eq:intW=0} before
taking the limit. 
In the correlator $\zcft^{{\rm reg}(\eps,L)}[W(z)]$ there
are $N=N_1+\cdots+N_r$ screening integrals, one along
each of the contours
$\gamma_{k,K}^{\eps,L}$. Let us denote
\be
  Q_k(\gamma_{k,K}^{\eps,L}) = \oint_{\gamma_{k,K}^{\eps,L}} 
  \frac{dx}{2\pi i} \, V_{\alpha_k}(x) ~.
\ee
The integration contour
$\gamma_{\rm ev}$ of $W(z)$ can be deformed to encircle
each of the $\gamma_{k,K}^{\eps,L}$. 
As in \erf{eq:A-A}, carrying out
the $z$ and the corresponding screening integration results 
in the insertion of a pair of fields
$A(y_1){-}A(y_0)$ at the endpoints
of the encircled contour $\gamma_{k,K}^{\eps,L}$.
The function $I(L)$ can thus be rewritten as follows,
\be\begin{array}{ll}
  I(L) &= \displaystyle
  \oint_{\gamma_{\rm ev}} \frac{dz}{2\pi i} \, \frac{1}{z{-}x}
  ~\langle \vec N| e^{-\Hh} ~\prod_{j,J} 
  Q_j(\gamma_{j,J}^{\eps,L}) ~ W(z) |0\rangle
  \\[5pt] &= \displaystyle
  \sum_{k,K} 
  \langle \vec N| e^{-\Hh} \hspace{-0.5em}\prod_{j,J \neq k,K}  \hspace{-1em}
  Q_j(\gamma_{j,J}^{\eps,L}) ~ \big(\, 
  A_k(y_1^{k,K}) -  A_k(y_0^{k,K}) \, \big)  |0\rangle ~.
\eear\labl{eq:app-IL-form}
Here 
$y_1^{k,K} = \gamma_{k,K}^{\eps,L}(L)$,
$y_0^{k,K} = \gamma_{k,K}^{\eps,L}(-L)$
and $A_k(y)$ denotes the field \erf{eq:A-form} with 
$V$ replaced by $V_{\alpha_k}$ and $f(y) = (y{-}x)^{-1}$.
The individual field insertions $A_k(y)$ are a sum of terms
of the form
\be
  ({\rm const}) \cdot \frac{1}{(y-x)^p} ~\partial^q
  \{V_{\alpha_k} W\}_r(y) ~.
\labl{eq:applim-1}
The state 
$|\psi\rangle = \partial^q \{V_{\alpha_k} W\}_r(0)|0\rangle$ 
is an element in the $U(1)^r$--highest weight
representation $\hcal_{\alpha_k}$ of charge $\alpha_k$ and
with highest weight vector $|\alpha_k\rangle$.
The space $\hcal_{\alpha_k}$ is spanned by the states
$J^{u_1}_{-m_1-1} \cdots J^{u_n}_{-m_n-1} |\alpha_k \rangle$,
for $m_i \ge 0$ and $u_i \in \lieh^*$. Thus the field
$\partial^q \{V_{\alpha_k} W\}_r(y)$ is itself a linear combination
of fields of the form
\be
  \phi(y) = (\partial^{m_1}J^{u_1} \cdots 
  \partial^{m_n} J^{u_n}V_{\alpha_k})(y) ~.
\labl{eq:applim-2}
In order to find an estimate for the integral $I(L)$ we will first
establish the formula
\bea\displaystyle
  \langle \vec N| e^{-\Hh} V_{q_1}(x_1) \cdots V_{q_n}(x_n) 
  \phi(y) |0\rangle 
  \\[5pt]\displaystyle
  =~ \frac{P(x_1,\dots,x_n  ,y)}{Q(x_1,\dots,x_n  ,y)}
  \langle \vec N| e^{-\Hh} V_{q_1}(x_1) \cdots V_{q_n}(x_n) 
  V_{\alpha_k}(y) |0\rangle 
\eear\labl{eq:app-phi-corr}
where $\sum_i q_i + \alpha_k = \vec N$, 
and $P$, $Q$ are polynomials, whose detailed form will not matter.
Take a field $\tilde\phi(y)$ which is also of the form
\erf{eq:applim-2} and consider the function $g(y)$ defined as
\be
  g(y) = 
  \langle \vec N| e^{-\Hh} ~ V_{q_1}(x_1) \cdots V_{q_n}(x_n) 
  ~ (\partial^m J^u \tilde\phi)(y) \, |0\rangle ~.
\ee
Using the definition of the normal ordered product we can
write
\bea\displaystyle
  g(y)=
  \oint_{\gamma_y} \frac{dz}{2\pi i} \, \frac{1}{z{-}y}
  \langle \vec N| e^{-\Hh} ~V_{q_1}(x_1) \cdots V_{q_n}(x_n) 
  ~\partial^m J^u(z) ~ \tilde\phi(y) \, |0\rangle 
  \\[5pt]\displaystyle
  = \Big( \oint_{\gamma_\infty}\frac{dz}{2\pi i} -
  \sum_{i=1}^n \oint_{\gamma_{x_i}}\frac{dz}{2\pi i} \Big)
  \frac{m!}{(z{-}y)^{m+1}}
  \langle \vec N| e^{-\Hh} ~ V_{q_1}(x_1) \cdots V_{q_n}(x_n) 
  ~ J^u(z) ~ \tilde\phi(y) \, |0\rangle
  \\[5pt]\displaystyle
  = \Big( \!\! -\frac{1}{g_s} \big( u, \partial^{m+1} \Wh(y) \big)
  - \sum_{i=1}^n \frac{m! ~ (q_i,u)}{(x_i{-}y)^{m+1}} \Big)
  \langle \vec N| e^{-\Hh} ~V_{q_1}(x_1) \cdots V_{q_n}(x_n) 
  ~\tilde\phi(y)  \,|0\rangle ~.
\eear\labl{eq:app-phi-rec}
In the first step, the integration contour $\gamma_y$ has been
deformed to encircle the points $x_i$ as well as the point $\infty$.
Also, by partial integration the $m$ derivatives $\partial/\partial z$ 
have been shifted
from $J^u(z)$ to the factor $(z{-}y)^{-1}$. The last step uses
the OPE \erf{eq:r-fb-block} as well as the fact that 
due to \erf{eq:JeH-com} the asymptotic
behaviour of the correlator in the second line of \erf{eq:app-phi-rec}
is given by
\bea\displaystyle
  \langle \vec N| e^{-\Hh} ~ V_{q_1}(x_1) \cdots V_{q_n}(x_n) 
  ~ J^u(z) ~ \tilde\phi(y)  \,|0\rangle
  \\[5pt]\displaystyle
  = -\frac{1}{g_s} \big( u,\Wh'(z) \big)
   ~ \langle \vec N| e^{-\Hh} ~ V_{q_1}(x_1) \cdots V_{q_n}(x_n) 
  ~ \tilde\phi(y) \, |0\rangle + O(z^{-1}) ~.
\eear\ee
Furthermore, it was used that for the contour integration around
infinity we have, for some polynomial $p(z)$,
\be
  \oint_{\gamma_\infty} \frac{dz}{2\pi i} \Big[
  \big( p(z) + O(z^{-1}) \big) 
  (-1)^m \frac{\partial^m}{\partial z^m} \frac{1}{z{-}y} \Big]
  = \partial^m p(y) ~.
\ee
Applying this procedure recursively allows us to strip
a field $\tilde\phi(y)$ of the form \erf{eq:applim-2} of all its 
components $\partial^{m_i} J^{u_i}$, establishing
equation \erf{eq:app-phi-corr}.

With the help of \erf{eq:app-phi-corr} we can now find
an estimate for a correlator involving screening integrals.
Consider the equalities
\bea\displaystyle
  \langle\vec N|e^{-\Hh} \hspace{-0.5em} \prod_{j,J \neq k,K} 
  \hspace{-0.8em} Q_j(\gamma_{j,J}^{\eps,L}) ~~
  \phi(y) \, |0\rangle
  \\[5pt]\displaystyle
  = \Big( \prod_{j,J \neq k,K} \oint_{\gamma_{j,J}}
  \frac{d\lambda_{j,J}}{2\pi i} \Big)
  \frac{P(\lambda_{1,1},\dots,\lambda_{r,N_r},y)
    }{Q(\lambda_{1,1},\dots,\lambda_{r,N_r},y)}
  \langle\vec N|e^{-\Hh} \hspace{-0.4em}\prod_{r,R \neq k,K} \hspace{-0.8em} 
    V_{\alpha_r}(\lambda_{r,R}) ~~V_{\alpha_k}(y) \,|0\rangle
  \\[5pt]\displaystyle
  = \bigg[\Big( \prod_{j,J \neq k,K} \oint_{\gamma_{j,J}}
  \frac{d\lambda_{j,J}}{2\pi i} \Big)
  \frac{\tilde P(\lambda_{1,1},\dots,\lambda_{r,N_r},y)
    }{\tilde Q(\lambda_{1,1},\dots,\lambda_{r,N_r},y)}
  e^{-\frac{1}{g_s}\sum_{r,R \neq k,K} W_r(\lambda_{r,R})}
  \bigg] e^{-\frac{1}{g_s} W_k(y)} ~.
\eear\labl{eq:app-phi-int}
Here $P,Q,\tilde P, \tilde Q$ are polynomials whose arguments
consist of $y$ as well as of all $\lambda$'s except for $\lambda_{k,K}$. 
The point $y$ depends on $L$ and 
is taken to have either of two values
\be
  y = y(L) = \gamma_{k,K}^{\eps,L}( \pm L) ~.
\ee
In the first step in \erf{eq:app-phi-int}
the screening integrals have been written out explicitly
and \erf{eq:app-phi-corr} was substituted. In the second
step the explicit form of the correlator, obtained
from \erf{eq:eHk-in-corr}, 
\erf{eq:r-fb-block} and \erf{eq:H-def}, has been
inserted and the rational part of the expression has been
absorbed by redefining the polynomials $P$ and $Q$ appropriately.

Denote by $F(L)$ the multiple integral inside the square
brackets in the last line of equation \erf{eq:app-phi-int}. For finite $L$ this
integral is finite, since the only singularities of the
integrand could
occur at coinciding integration variables $\lambda_{i,I} = \lambda_{j,J}$
which is impossible by construction of the regularised
contours $\gamma_{i,I}^{\eps,L}$.
Furthermore,
because of the asymptotic behaviour of the contour
$\gamma(x)$ imposed at the end of section \ref{sec:contour}, the 
integrand of $F(L)$ receives an exponential damping as its
arguments approach infinity. We can conclude that the
$L\rightarrow\infty$ limit of $F(L)$ is well defined,
\be
  \lim_{L\rightarrow \infty} F(L) = C ~~,
\ee
for some constant $C$. For the correlator in the first line of
\erf{eq:app-phi-int} this implies that
\be
  \lim_{L\rightarrow \infty} 
  \langle\vec N|e^{-\Hh} \hspace{-0.5em} \prod_{j,J \neq k,K} 
  \hspace{-0.8em} Q_j(\gamma_{j,J}^{\eps,L}) ~~
  \phi(y) \, |0\rangle
  ~=~ \lim_{L\rightarrow \infty}  F(L) e^{-\frac{1}{g_s} W_k(y(L))} ~=~ 0
  ~~.
\ee
Since from \erf{eq:app-IL-form} we know that $I(L)$ can be written as
a linear combination of terms of the form \erf{eq:app-phi-int}, the
above equation proves that indeed $\lim_{L\rightarrow\infty} I(L)=0$,
as claimed.

\end{document}